\documentclass{ar-1col}

\usepackage[comma]{natbib}
\usepackage{url,amsmath}

\def\aap{{\em Astr.~Astrophys.}}

\def\apj{{\em Astrophys.~J.}}
\def\apjl{{\em Astrophys.~J.~Lett.}}
\def\apjs{{\em Astrophys.~J.~Suppl.}}

\def\araa{{\em Ann.~Rev.~Astr.~Astrophys.}}

\def\mnras{{\em Mon.~Not.~R.~astr.~Soc.}}
\def\nat{{\em Nature}}

\def\physrep{{\em Physics Reports}}

\def\prd{{\em Phys.~Rev.~D}1}

\def\ssr{{\em Space Sci.~Rev.}}

\def\cm{{\rm\thinspace cm}}

\def\erg{{\rm\thinspace erg}}
\def\eV{{\rm\thinspace eV}}
\def\g{{\rm\thinspace g}}

\def\K{{\rm\thinspace K}}
\def\keV{{\rm\thinspace keV}}
\def\km{{\rm\thinspace km}}

\def\Msun{\hbox{$\rm\thinspace M_{\odot}$}}

\def\s{{\rm\thinspace s}}

\def\Hz{{\rm\thinspace Hz}}

\def\ergpcmsqps{\hbox{$\erg\cm^{-2}\s^{-1}\,$}}

\def\ergps{\hbox{$\erg\s^{-1}\,$}}

\def\gps{\hbox{$\g\s^{-1}\,$}}
\def\kmps{\hbox{$\km\s^{-1}\,$}}

\def\h18{\hbox{H1821$+$643\,}}

\jname{Xxxx. Xxx. Xxx. Xxx.}
\jvol{AA}
\jyear{YYYY}
\doi{10.1146/((please add article doi))}

\begin{document}

\markboth{C.~S.~Reynolds}{Black Hole Spin}

\title{Observational Constraints on Black Hole Spin}

\author{Christopher S. Reynolds $^1$
\affil{$^1$Institute of Astronomy, University of Cambridge, Madingley Road, Cambridge, CB3 OHA, United Kingdom; email: csr12@ast.cam.ac.uk}
}
\begin{abstract}
The spin of a black hole is an important quantity to study, providing a window into the processes by which a black hole was born and grew.  Further, spin can be a potent energy source for powering relativistic jets and energetic particle acceleration.  In this review, I describe the techniques currently used to detect and measure the spins of black holes.  It is shown that:

\parbox{10cm}{
\begin{itemize}
\item Two well understood techniques, X-ray reflection spectroscopy and thermal continuum fitting, can be used to measure the spins of black holes that are accreting at moderate rates.  There is a rich set of other electromagnetic techniques allowing us to extend spin measurements to lower accretion rates.
\item Many accreting supermassive black holes are found to be rapidly-spinning, although a population of more slowly spinning black holes emerges at masses above $M>3\times 10^7\Msun$ as expected from recent structure formation models.
\item Many accreting stellar-mass black holes in X-ray binary systems are rapidly spinning and must have been born in this state. 
\item The advent of gravitational wave astronomy has enabled the detection of spin effects in merging binary black holes.  Most of the pre-merger black holes are found to be slowly spinning, a notable exception being an object that may itself be a merger product. 
\item The stark difference in spins between the black hole X-ray binary and the binary black hole populations shows that there is a diversity of formation mechanisms.  
\end{itemize}
}
Given the array of new electromagnetic and gravitational wave capabilities currently being planned, the future of black hole spin studies is bright. 
\end{abstract}

\maketitle

\tableofcontents

\section{INTRODUCTION}\label{intro}

One of the most remarkable predictions of Einstein's Theory of General Relativity (GR) is the existence of black holes, objects in which the total victory of gravity over the other forces of nature has created a space-time singularity cloaked by an event horizon that marks the point of no return for in-falling matter or light. There are two great ironies that one immediately confronts when considering black holes.  Firstly, while they are themselves the darkest objects in the Universe, we find that the most luminous sources in the Universe are powered by the accretion of matter into a black hole. Indeed, today's standard models for galaxy formation and evolution recognise that the energy released building central supermassive black holes (SMBHs) is sufficient to profoundly impact their host galaxy \citep{fabian:12a}.  Secondly, the processes by which black holes form in the cores of collapsing massive stars or the centers of protogalaxies, the physics of accretion onto black holes via radiation-dominated, turbulent plasma disks \citep{shakura:73a,balbus:91a}, and the physical processes by which SMBHs feedback on galactic scales \citep{fabian:12a} are amongst some of the most complex and actively researched topics within astrophysics.  And yet, GR tells us that black holes are fundamentally the simplest macroscopic objects in the Universe. The no-hair theorem of GR 
states that, when an isolated black hole is formed, the properties of the final object are fully characterized by its mass, angular momentum, and electrical charge, i.e. the minimum set of quantities compatible with the conservation laws of classical physics. In fact, in any realistic setting, a charged black hole will be neutralized by discharge to its surroundings, so astrophysical black holes are defined by only mass and angular momentum (hereafter referred to as simply spin).\begin{marginnote}[]
\entry{GR}{Theory of General Relativity describing gravity in terms of spacetime curvature.}
\entry{Singularity}{Location at the black hole's center where the GR must break down.}
\entry{Event Horizon}{Point of no return within which all matter and light is drawn in.}
\entry{No hair theorem}{Statement that black holes are specified completely by just mass, spin, and charge}
\entry{SMBH}{Supermassive black holes, with masses $10^6-10^{10}\Msun$ and found in galactic centers.}
\end{marginnote}

Black hole spin has a significant effect on astrophysical phenomena. The efficiency with which accreting matter is converted into electromagnetic radiation can vary by almost an order of magnitude depending upon whether the matter accretes onto a slowly or rapidly rotating black hole, and whether it accretes in a prograde or retrograde sense \citep{bardeen:72a}.  Similarly, the efficacy with which black holes produce relativistic jets carrying significant kinetic luminosity is likely tied to whether they are rapidly spinning or not. And spin-induced relativistic precession effects around black holes can  spray these collimated jets across a wide range of solid angles thereby increasing their coupling to the surroundings.

Spin is also a fossil-remnant of how the black hole formed and grew. For the stellar-mass black holes found in Galactic X-ray binary systems, there is little doubt that the black holes we see today were formed via the core-collapse of a massive star.  In these cases, the spin of these black holes is a window on the processes that occurred during that stellar core collapse. For every other class of black hole known --- the SMBHs in galactic cores, the binary black holes detected by gravitational wave observatories, and the accreting black holes in dense stellar clusters --- their history is much less certain, and spin can help distinguish different scenerios for their growth.\begin{marginnote}[]
\entry{X-ray binary}{A system consisting of a star that is transferring mass to an orbiting relativistic compact object, either a neutron star or a black hole.}
\end{marginnote}

Any rigorous understanding of these issues must start from actual measurements of black hole spin. This is not an easy task. The mass of a black hole can be determined from its far-field gravitational influence, e.g. via the orbit of the companion star in the case of a black hole X-ray binary, or the motion of broad-line region clouds in an active galactic nucleus (AGN). But black hole spin is an inherently relativistic phenomenon and observations must probe the strong gravity region close to the black hole if one is to determine spin. This has been a long road for ``traditional'' electromagnetic (EM) astrophysical studies of accreting black holes; our ability to measure spin is intertwined with our understanding of accretion disk physics, and the two fields of study have had to undergo a long and sometimes painful co-evolution.  Progress has been good, however, and there are now multiple techniques by which the spin of an accreting black hole can be probed.
\begin{marginnote}[]
\entry{AGN}{Active galactic nucleus, an actively accreting supermassive black hole at the center of a galaxy.}
\entry{EM}{Electromagnetic radiation, the traditional carrier of astronomical information.}
\end{marginnote}

The advent of gravitational wave (GW) astronomy has opened a new window on black hole spin via the study of merging binary black holes (BBHs). The GW signals are free from ``astrophysical'' complexities (i.e. gas/plasma physics) but, in these early stages of GW astronomy, the data quality enable only  limited spin constraints.  Even so, GW studies have revealed new populations of stellar-mass black holes, and the spin measurements are important for constraining formation and evolution scenarios.
\begin{marginnote}[]
\entry{GW}{Gravitational waves, ripples in the curvature of spacetime that propagate at the speed of light.}
\entry{BBH}{Binary black hole, two black holes orbiting their common center of mass.}
\end{marginnote}

In this review, we survey the current state-of-the-art in measurements of black hole spin.  There have been many excellent reviews of EM methods for measuring black hole spin \citep[e.g. ][]{miller:07a,brenneman:13a,miller:15a,middleton:16a,bambi:18a} as well as several by the author \citep{reynolds:13a,reynolds:14a,reynolds:19a}.  In addition to updating these previous works, the current article benefits from the tremendous progress in GW astronomy, providing one of the first opportunities for a spin-based comparison of the black hole populations studied by GW astronomy and traditional EM astronomy. 

\section{THE PHYSICS OF A SPINNING BLACK HOLE}\label{physics}

Within a GR-framework, the exact description of an isolated, spinning, and uncharged black hole is given by the Kerr metric \citep{kerr:63a,mtw:73a}.  The structure of the spacetime around a single black hole is solely characterized by the black hole's mass $M$ and angular momentum $J$.  It is common to define a dimensionless spin parameter $a\equiv Jc/GM^2$, and almost all of our discussion of black hole spin will be in terms of this parameter. \begin{marginnote}[]
\entry{$a$}{Dimensionless spin parameter of the black hole, spanning the range from $-1$ to $+1$.}
\end{marginnote}
Using the Boyer-Lindquist coordinates ${t,r,\theta,\phi}$ (which are the most straightforward generalization of spherical polar coordinates to this spacetime structure), the line element of the Kerr metric reads
\begin{eqnarray}
  ds^2&=&-\left(1-\frac{2Mr}{\Sigma}\right)dt^2 -
  \frac{4aM^2r\sin^2\theta}{\Sigma}dt\,d\phi +
  \frac{\Sigma}{\Delta}dr^2\\\nonumber
&+&\Sigma\,d\theta^2+\left(r^2+a^2M^2+
\frac{2a^2M^3r\sin^2\theta}{\Sigma}\right)\sin^2\theta\,d\phi^2,
\end{eqnarray}
where $\Delta=r^2-2Mr+a^2M^2$, $\Sigma = r^2+a^2M^2\cos^2\theta$, and just here we have adopted the traditional relativists approach of setting $G=c=1$. \begin{marginnote}[]
\entry{Kerr metric}{Description of the spacetime structure around an isolated spinning black hole.}
\end{marginnote}

Detailed explorations of this metric are presented in many standard texts \cite[e.g.][]{mtw:73a}.  Here, we focus on just those main properties that are relevant for the astrophysically-oriented discussions that follow, and we do so at a level that does not require advanced training in GR. At the very center of the black hole, there is a location where the spacetime curvature diverges and we have a spacetime singularity (that formally possesses a ring-like topology).  But, provided that $-1\le a\le 1$, this singularity is cloaked by an event horizon at $r_{\rm evt}=\left(1+\sqrt{1-a^2}\right)r_g$ (i.e. the outer root of $\Delta=0$), where $r_g=GM/c^2$ is the standard definition of the gravitational radius. The event horizon marks the point of no return for any infalling matter or light, and no signal emitted from $r<r_{\rm evt}$ may ever propagate outwards.  

If we were to permit $|a|>1$, there would be no event horizon and the spacetime singularity would be naked for all to see, a situation deemed to be forbidden by the cosmic censorship hypothesis. In this sense, $|a|=1$ (known as the Kerr bound) is the ``rotational break-up'' limit for a rotating black hole. In fact, provided that the initial collapse respects the Kerr bound, it is impossible to spin-up a black hole beyond this limit through any normal accretion processes.  We note in passing an interesting thread of String Theory inspired research which suggests that true singularities can be avoided at the cores of superspinning ($|a|>1$) compact objects.  In such objects, the pathological singularity at $r=0$ in the Kerr solution is replaced by a spherical domain wall comprised of strings and D-branes \citep{gimon:09a}.  Superspinars would have characteristics rather distinct from black holes \citep{schee:13a}, including the near 100\% conversion into radiation of the gravitational energy of infalling matter. The observational discovery of a super-spinner would be profoundly important for fundamental physics,

Spin introduces a qualitatively new feature into the physics --- inertial frames of reference are dragged into rotation around the black hole. This has a plethora of astrophysical consequences. If a particle orbits the black hole in a plane that it not aligned with the black hole spin (i.e. an orbit does not lie in the $\theta=\pi/2$ plane), the orbital plane will undergo Lens-Thirring precession with frequency $\Omega_{\rm LT}=2a(GM)^2/c^5r^3$.  But even particles on circular orbits in the symmetry plane of the black hole spin are affected by frame-dragging.  In black hole spacetimes, circular orbits are only stable outside of a critical radius known as the innermost stable circular orbit (ISCO).  For a non-rotating black hole the ISCO is at $r_{\rm isco}=6r_g$, but black hole spin has a significant effect on the location of the ISCO. If a particle orbits in the same sense as the black hole spin (i.e. prograde), the action of frame-dragging can stabilize otherwise unstable orbits, moving the ISCO closer to the black hole (with $r_{\rm isco}=r_g$ for $a=1$).  Conversely, retrograde orbiting particles tend to be destabilized by frame-dragging, moving the ISCO away from the black hole (with $r_{\rm isco}=9r_g$ for $a=-1$). In the general case, the ISCO is given by,\begin{marginnote}[]
\entry{Lens-Thirring precession}{The rotation of an orbital plane around the spin axis of a black hole.}
\entry{ISCO}{Innermost stable circular orbit, the closest stable circular orbit to the black hole.}
\end{marginnote}
\begin{eqnarray}
r_{\rm isco}&=&\left(3+Z_2\mp \left[(3-Z_1)(3+Z_1+2Z_2) \right]^{1/2} \right)\,r_g,
\end{eqnarray}
\citep{bardeen:72a} where we consider orbits restricted to the $\theta=\pi/2$ plane, the $\mp$ sign is for particles in prograde/retrograde orbits, respectively, and we have defined,
\begin{eqnarray}
Z_1&=&1+\left(1-a^2\right)^{1/3}\left[\left(1+a\right)^{1/3}+\left(1-a\right)^{1/3}\right],\\
Z_2&=&\left(3a^2+Z_1^2\right)^{1/2}.
\end{eqnarray}
As discussed in Section~\ref{disks}, accreting matter flowing into the black hole via a geometrically-thin disks is expected to start its final plunge into the black hole once it crosses the ISCO. 

A particle that is spiralling on almost circular orbits must lose energy in order to migrate inwards. If the particle can be kept on circular orbits even after it crosses the ISCO (due to radial pressure forces in a geometrically-thick accretion disk, or electromagnetic forces associated with an ordered magnetic field) then its total energy starts to again increase.   The marginally-bound orbit, where the particle has the same total energy as it would if it were at rest at infinity, is spin-dependent and is at a radius\begin{marginnote}[]
\entry{Marginally bound orbit}{Orbit close to a black hole on which a particle has the same binding energy as it would at rest at infinity.}
\end{marginnote}
\begin{equation}
r_{\rm mb}=\left(2\mp a+2\sqrt{1\mp a}\right)\,r_g,
\end{equation}
\citep{bardeen:72a} where, again, we have restricted to orbits within the $\theta=\pi/2$ plane and the $\mp$ sign is for particles in prograde/retrograde orbits.  On energetic grounds, this radius is expected to define the inner edge of the Keplerian part of any accretion disk, even ones that are geometrically-thick with significant radial pressure gradients.

Inside of the marginally-bound orbit is the special location where photons can orbit the black hole, the photon circular orbit.  This radius is also spin-dependent and given by
\begin{equation}
r_{\rm ph}=2\left(1+\cos\left[{2\over 3}\cos^{-1}(\mp a)\right]\right)\,r_g,
\end{equation}
\citep{bardeen:72a}.   Again, the $\mp$ sign corresponds to prograde/retrograde orbits respectively.  Inside of this radius, circular orbits do not exist.  In cases where a black hole is surrounded by an optically-thin accretion disk, the photon circular orbit shows up in direct images as a bright ring (see Section~\ref{imaging}).   \begin{marginnote}[]
\entry{Photon circular orbit}{Special orbit close to a black hole on which a photon can go in a closed circle.}
\end{marginnote}

Very close to a spinning black hole, frame dragging effects become extreme. Interior to the surface defined by $\Sigma=2Mr$, known as the static limit, frame-dragging forces any matter or even photons to rotate in the same sense as the black hole as seen by a distant observer. \cite{penrose:69a} showed through a series of thought experiments that physical processes (e.g. particle-particle scattering) interior to the static limit can tap into and extract the spin energy of the black hole --- this gives the name to this region, the ergosphere (ergon meaning `work' in Ancient Greek). Considerations based on black hole thermodynamics tells us that the (extractable) energy associated with black hole rotation is
\begin{equation}
E_{\rm spin}=\left[1-\frac{1}{2}\left(\left[1+\sqrt{1-a^2}\,\right]^2+a^2\right)^{1/2}\right]Mc^2,
\end{equation}
 \citep{mtw:73a}.  For an extremal black hole ($|a|=1$) this gives $E_{\rm spin}=(1-1/\sqrt{2})Mc^2$ which is a colossal 29\% of the rest mass energy of the black hole. Thus we come to the important realization that black hole spin can be an astrophysically important energy source. The processes envisaged in Penrose's original work require quite violent particle interactions within the ergosphere \citep{bardeen:72a}, and may produce high-energy particle acceleration and $\gamma$-ray emission \citep{williams:95a}.  Most of the attention by the community, however, has focused on a magnetic version of the Penrose process whereby magnetic fields threading either the event horizon \citep{blandford:77a} or orbiting matter within the ergosphere \citep{gammie:99a} can tap into the black hole spin. Indeed, the standard paradigm for the relativistic jets seen from many black hole systems is that they are powered by the magnetic extraction of black hole spin energy \citep[][also see Section~\ref{jet}]{begelman:84a}.

\begin{figure}
\includegraphics[width=4in]{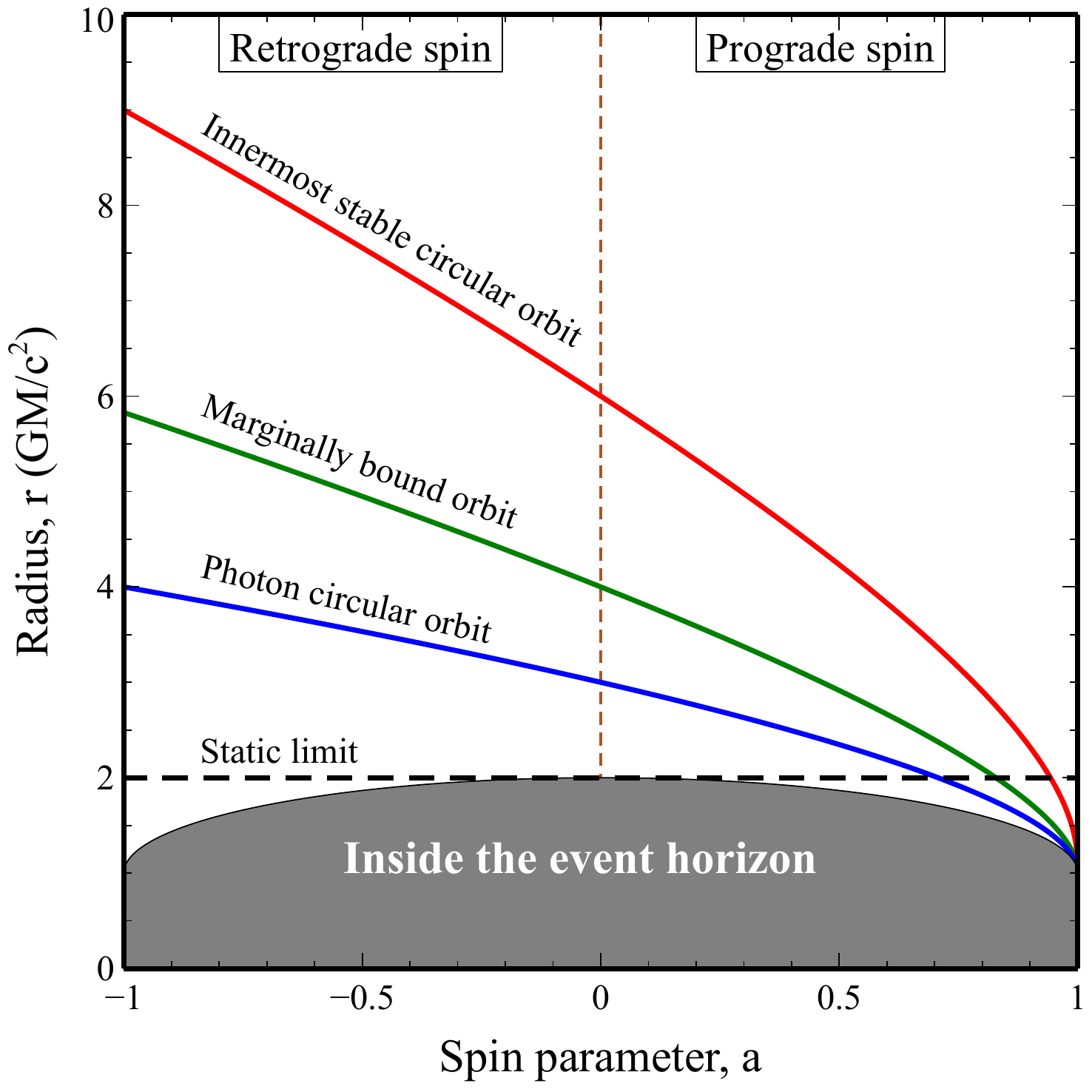}
\caption{Boyer-Lindquist radius of special orbits in the equatorial plane a Kerr black hole as a function of dimensionless spin parameter $a$.  See Section~\ref{physics} for detailed discussion.}
\label{fig:bhradii}
\end{figure}

We summarize the spin-dependence of these special locations in Figure~\ref{fig:bhradii}.  We note that the ISCO, the marginally bound orbit, the photon circular orbit and the event horizon all seem to converge on $r=GM/c^2$ as $a\rightarrow 1$.  This is in fact due to the singular nature of the Boyer-Lindquist radial coordinate in the extremal ($a=1$) Kerr metric.  A careful consideration of proper distance reveals that these four locations remain distinct and separated by finite proper distance even as $a\rightarrow 1$.

Our discussion so far has focused on the properties of a single black hole but, with the advent of GW astronomy, it is important to consider merging BBHs.  GR is a highly non-linear theory and so we cannot simply superpose two Kerr solutions in order to describe a BBH. Indeed, analytic solutions do not exist for such systems and, in general, the full time-dependent field equations of GR must be solved numerically in order to completely describe the inspiral and merger of two black holes. Even if the black holes are on circular orbits, the geometry of the binary can be complex with the two black hole spins in general being mis-aligned both to each other and to the total orbital angular momentum. The physics of how these spins imprint on observable gravitational waves, and the possible simplifications used to extract this from real signals, is deferred to Section~\ref{gwa}.

\section{MEASURING BLACK HOLE SPIN IN ACCRETING SYSTEMS}\label{accreting}

Ultimately, spin measurements require us to probe frame-dragging effects that only become significant close to the black hole.  Until the recent advent of GW astronomy, spin measurements focused exclusively on accreting black holes, requiring us to characterize and measure the influence of the spin on the accretion processes.  Black hole accretion is itself a complex and rapidly advancing topic, so observational studies of spin must advance hand-in-hand with those of black hole accretion in general.  The result can be a rather bewildering literature that includes apparent disagreements and controversies. We attempt to capture the current state-of-play in as orderly a manner as possible.

\subsection{Brief primer on black hole accretion physics}\label{disks}

There are many excellent reviews on the modern theory of black hole accretion \citep[e.g. ][]{davis:20a}; here we give a very brief overview, focusing on issues that have direct bearing on the topic of black hole spin. 

Due to the angular momentum that infalling matter will inevitably possess, accreting matter will form a rotationally supported disk around the black hole. To a large extent, the physics of accretion is then dictated by how the matter can rid itself of angular momentum. The standard paradigm is that the accretion flow becomes turbulent due to the action of the magnetorotational instability \citep[MRI; ][]{balbus:91a}, and that correlated magnetic forces within the magnetohydrodynamic (MHD) turbulence lead to the outward transport of angular momentum. Angular momentum can also be directly removed from the accretion disk by a magnetized wind \citep{blandford:82a}. As a result, matter gradually spirals into the black hole in approximately circular orbits until, at some location, it undergoes a rapid plunge into the black hole itself. \begin{marginnote}[]
\entry{MHD}{Magneto-hydrodynamics, the formal description of the dynamics of a magnetized fluid.}
\entry{MRI}{Magnetorotational instability, responsible for turbulence and angular momentum transport in accretion disks. }
\end{marginnote}

While much of the physics of disk accretion relates to angular momentum, many of the observables (including those related to the details of how spin affects the disk) depend upon the thermodynamics of the accretion disk.  This, in turn, depends upon the accretion rate.
As a fiducial standard, we define the Eddington accretion rate,
\begin{equation}
\dot{M}_{\rm Edd}=\frac{L_{\rm Edd}}{\eta c^2},
\end{equation}
where $\eta$ is the radiative efficiency of the accretion flow, and $L_{\rm Edd}=4\pi GMm_pc/\sigma_{\rm T}$ is the standard Eddington luminosity which (when emitted isotropically) would produce a radiative force on fully ionized hydrogen that balances gravity. \begin{marginnote}[]
\entry{Eddington luminosity}{Critical luminosity where radiation force balances gravity for fully ionized hydrogen.}
\entry{$\dot{M}_{\rm Edd}$}{Eddington accretion rate, the mass flow rate that gives the Eddington luminosity}
\end{marginnote}

For accretion rates in the range $\dot{M}\sim 0.01-0.3\dot{M}_{\rm Edd}$, the accretion disk can cool efficiently and is expected to become geometrically-thin in the sense that its vertical scale height $h$ is much less than the distance from the black hole $r$.  In our standard picture for such disks \citep{shakura:73a}, radial pressure gradients and radial magnetic forces are unimportant and matter follows almost test-particle orbits in the black hole potential.  Once the accreting matter reaches the ISCO, it will plunge into the black hole on orbits that conserve energy and angular momentum. Thus, such accretion disks possess a distinct transition at the ISCO, from a turbulent, dissipative, circulating flow outside the ISCO to a laminar, non-dissipative, plunging flow within.  As a practical matter, the ISCO defines the ``inner edge'' to the region of the accretion disk responsible for observed emission.  Since the ISCO has a strong spin-dependence, this induces a spin-dependence to four key observables from the inner disk, (i) the maximum observed gravitational redshift, (ii) the maximum observed disk temperature, (iii) the maximum temporal frequencies observed, and (iv) the overall radiative efficiency $\eta$. The majority of spin-measurements published to date are based, ultimately, on this hypothesis that the ISCO effectively truncates the observable disk \citep{reynolds:08a,shafee:08b}.\begin{marginnote}[]
\entry{h}{Disk thickness, the scale-height of the disk in the vertical direction.  }
\end{marginnote}

As one considers higher accretion rates, we eventually expect a breakdown of the thin-disk condition $h\ll r$. According to the \cite{shakura:73a} model, radiation-pressure dominates over gas pressure within the inner disk once the mass accretion rate exceeds $\dot{M}\sim 10^{-2}\dot{M}_{\rm Edd}$ for AGN, and  $\dot{M}\sim 10^{-1}\dot{M}_{\rm Edd}$ for accreting stellar mass black holes.  The radiation-dominated portion of the accretion disk is then expected to have  a scale-height approximately given by 
\begin{equation}\label{eqn:h}
h\approx\frac{3}{2\eta}\left(\frac{\dot{M}}{\dot{M}_{\rm Edd}}\right)J(r)(1-f_X)\,r_g,
\end{equation}
where $J(r)=\left(1-\sqrt{r_{\rm isco}/r}\right)$ and so describes a disk that has constant-$h$ well beyond the ISCO but tappers down to $h=0$ at the ISCO.  Here, we have allowed for the possibility that a fraction $f_X$ of the energy is removed magnetically from the optically-thick accretion disk to be radiated in an optically-thin X-ray emitting corona \citep{svensson:94a}. Since $\eta\sim 0.1$ (see Section~\ref{efficiency}) we see that there is a potential breakdown of the thin-disk assumption in the inner disk once accretion rates exceed $\sim0.3\dot{M}_{\rm Edd}$, although this can be delayed to significantly higher accretion rates for disks with very active coronae ($f_X> 0.5$). \begin{marginnote}[]
\entry{$\eta$}{Radiative efficiency of the accretion flow, the fraction of the rest-mass energy of the infalling matter converted to EM radiation.}
\end{marginnote}Once radiation pressure significantly thickens the disk, the ISCO ceases to have any special role to play and the disk can in principle extend down to the marginally-bound order before plunging into the black hole. More importantly from a practical point of view, however, is that such luminous systems will likely have strong radiation-driven outflows. For genuinely super-Eddington systems ($\dot{M}\gg\dot{M}_{\rm Edd}$), scattering and reprocessing by these outflows will have a significant effect on observables \citep[e.g. ][]{kara:16a}, making signatures of spin very difficult to discern.  

For low accretion rates ($\dot{M}<10^{-2}\dot{M}_{\rm Edd}$), it is widely believed that the inner accretion flow becomes optically-thin and radiatively-inefficient \citep{rees:82a,narayan:95a}.  Energy dissipated in the MHD turbulence now remains as internal energy and is advected radially inwards rather than being radiated away.  The disk heats up to the virial temperature, and becomes geometrically-thick ($h/r\sim 0.5-1$). As with the high accretion rate case, the transition from the orbiting flow to the inward plunging flow is expected to be smeared out in such geometrically-thick flows and the ISCO will cease to have an identifiable imprint on the dynamics. There are other means of accessing spin information in such systems, however. These geometrically-thick disks are thought to be ideal for supporting large scale poloidal magnetic fields that can thread a central black hole and tap into the spin energy, launching relativistic jets.  The use of  jet power as a probe of black hole spin is explored in Section~\ref{jet}. Geometrically-thick accretion flows can also undergo large-scale precession due to Lens-Thirring torques, imprinting spin-dependent quasi-periodic oscillations into the observed flux (Section~\ref{qpos}). And finally, direct imaging of these optically-thin accretion flows allows us to characterize the photon circular orbit (Section~\ref{imaging}).

Having described the various ways that spin can affect accretion flows around single black holes, we now proceed to a detailed discussion of methods that have been employed to extract spin from real data.  

\subsection{Spin from X-ray reflection spectroscopy}\label{reflection}

The X-ray reflection method is applicable to systems in the ``Goldilocks zone'' of geometrically-thin/optically-thick accretion disks and measures spin through the gravitational redshift of spectral features from close to the ISCO. Thus, it is applicable to systems with accretion rates of $\dot{M}\sim 0.01-0.3\dot{M}_{\rm Edd}$, although this range may be extended up to almost Eddington-limited sources if substantial energy is transported into the corona \citep[eqn.~\ref{eqn:h}; ][]{svensson:94a}. Many of the classical Seyfert galaxies, moderate luminosity quasars, and black hole X-ray binaries in their luminous hard state are thought to be in this accretion rate range.

\subsubsection{Physics of X-ray reflection from black hole accretion disks}
\begin{figure*}
\hbox{
\hspace{-4cm}
\includegraphics[width=0.49\textwidth]{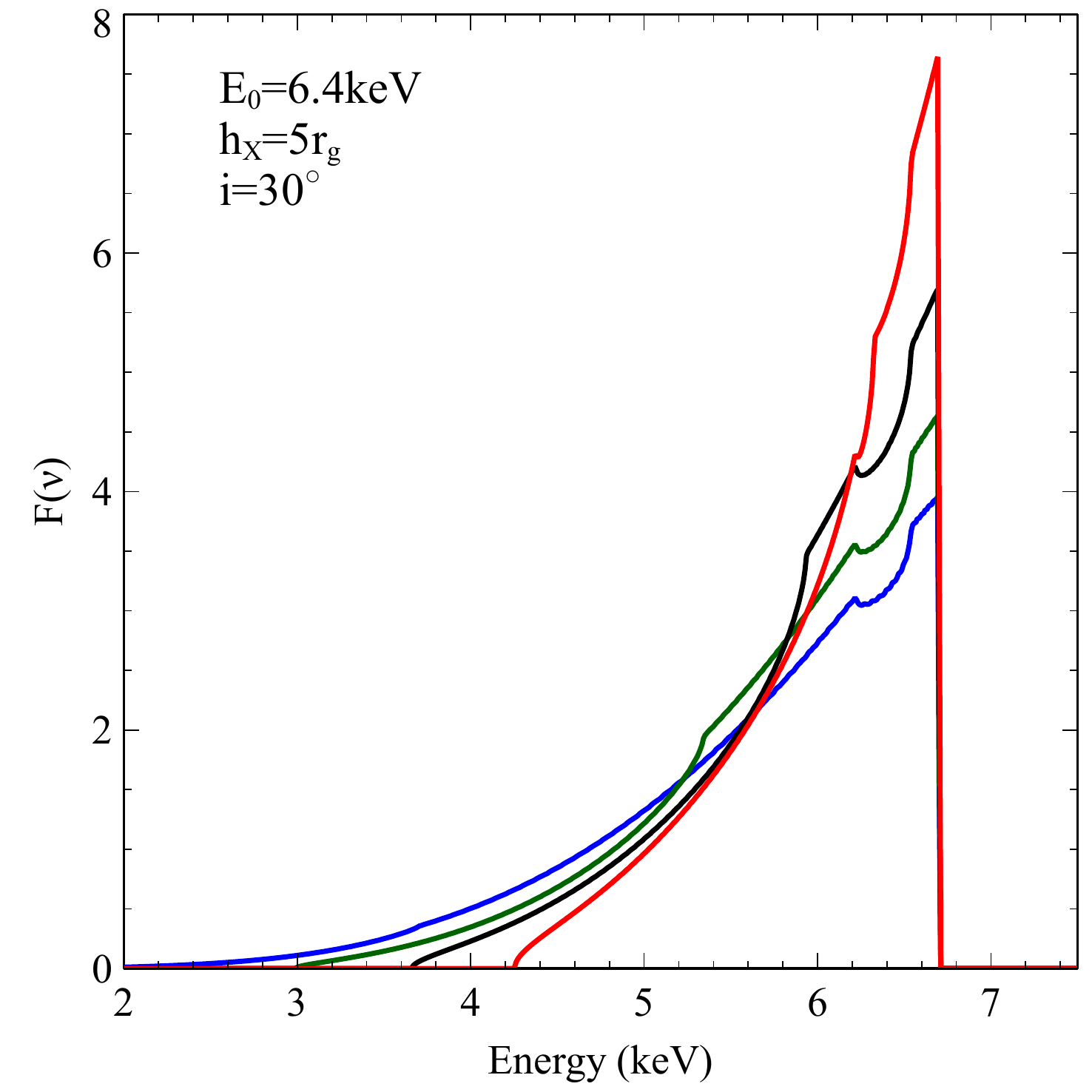}
\hspace{-6.5cm}
\includegraphics[width=0.49\textwidth]{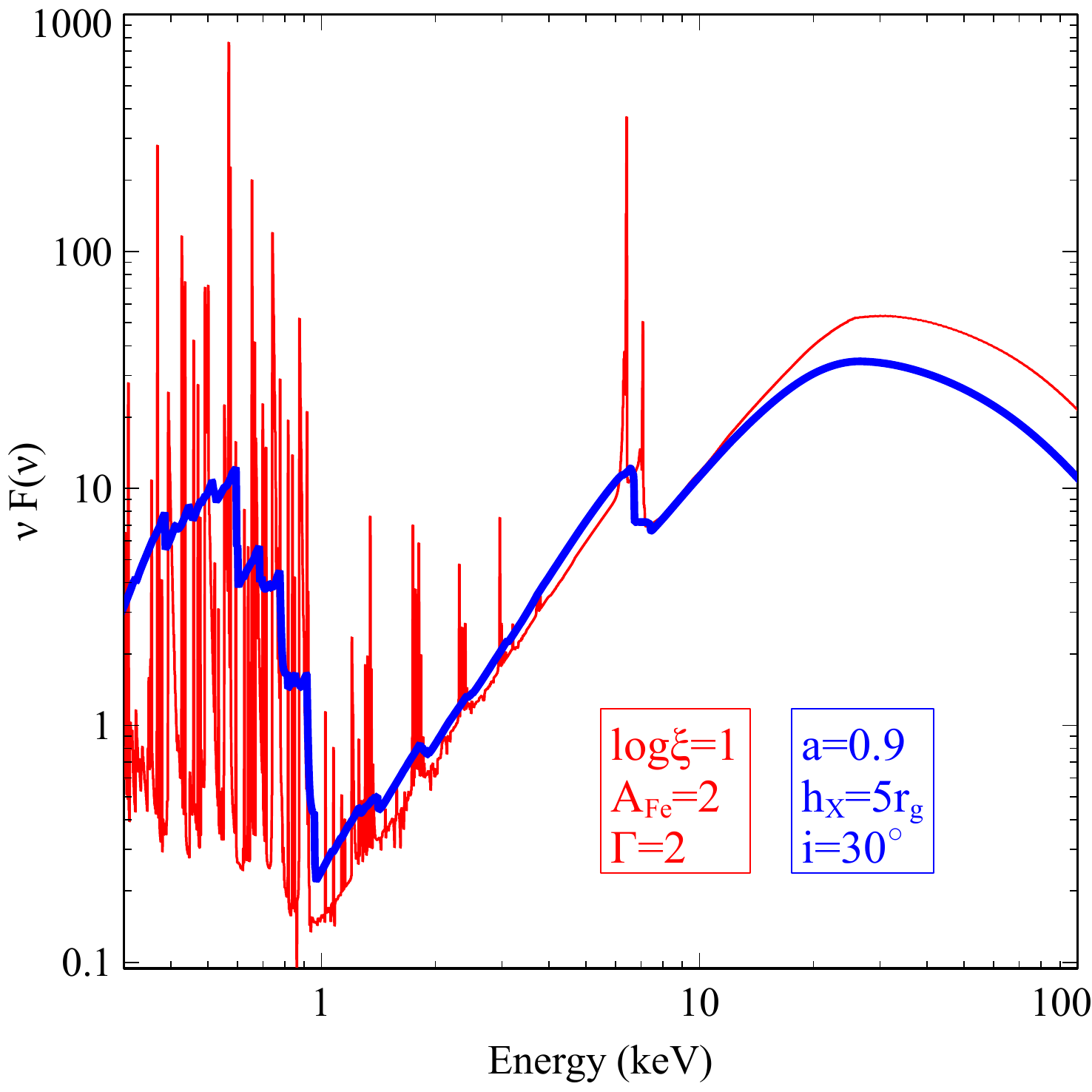}
}
\caption{Illustration of the two key ingredients underpinning the X-ray reflection method for spin measurements. {\it Left panel : }Illustrative profiles of a single emission line from the surface of a geometrically-thin accretion disk around a Kerr black hole.  The line is broadened and skewed through the combined action of relativistic Dopper shifts, relativistic beaming, and gravitational redshifts.  We assume a rest-frame line energy of $E_0=6.4\keV$ (i.e. the energy of neutral iron-$K\alpha$ fluroescence), a viewing inclination for the disk of $i=30^\circ$ away from face-on, and a line emissivity as a function of radius corresponding to excitation by point source located on the black hole spin axis at a distance of $h_X=5r_g$ (i.e. a lamppost geometry).  Four spins are shown; $a=-0.9$ (red), $a=0$ (black), $a=0.5$ (green) and $a=0.9$ (blue).  {\it Right panel : }Example rest-frame (red) and observed (blue) X-ray reflection spectrum.  The rest-frame spectrum assumes that the surface of the disk (electron density $n_e$) is irradiated by an X-ray power-law with photon index $\Gamma=2$ and ionizing flux $F_i$ resulting in an ionization parameter $\xi=4\pi F_i/n_e=10\ergps\cm$.  We assume that the iron abundance is twice the solar value ($A_{\rm Fe}=2$). This spectrum is dominated soft X-ray radiative recombination lines, powerful iron K-shell lines in the 6.4--6.97\,keV band, and a Compton reflection hump peaking at 20--30\.keV.  The relativistic blurring in the observed spectrum matches the $a=0.9$ case (blue) from the left panel. Both plots are constructed using the {\tt relxill} package \citep{garcia:14a}.}
\label{fig:reflectionlines}
\end{figure*}

In AGN and some states of black hole X-ray binaries, a significant amount of luminosity is in a hard X-ray power-law tail that extends to $\sim 100\keV$. This is initially very surprising --- these optically-thick accretion disks should produce quasi-blackbody spectra with temperatures of $kT\sim 0.01-0.1\keV$ for AGN and $kT\sim 1\keV$ for stellar-mass black holes (see Section~\ref{continuum}).   From the early days of black hole studies, it was realized that there must be a hot/energetic corona above the accretion disk that produces the hard X-rays via inverse Compton scattering of the disk thermal emission \citep{thorne:75a}. The coronal X-rays must then irradiate and partially photoionize the photosphere of the optically-thick accretion disk. In response, the surface layers of the accretion disk will emit a spectrum rich in X-ray fluorescent and radiative-recombination emission lines superposed on a free-free and Compton-scattered continuum \citep{fabian:89a,george:91a}.  This is traditionally referred to as the X-ray reflection spectrum and is characterized by a forest of soft X-ray lines, strong K-shell emission lines of iron (in the 6.4--6.97\,keV band depending upon ionization state) and a Compton scattered continuum that is shaped into a ``hump'' via the photoelectric edge of iron on the low side and Compton down-scattering at high energy (Fig.~\ref{fig:reflectionlines}). \begin{marginnote}[]
\entry{Corona}{Hot ($10^9\K$) plasma close to the black hole that produces observed hard X-ray emission.}
\end{marginnote}

The observed X-ray reflection spectrum is distorted by the Doppler effect associated with the orbital motion of the matter in the accretion disk and the gravitational redshift of the black hole potential. Both of these effects become stronger as one approaches the black hole. Emission lines develop a characteristic skewed profile with a sharp blue-shifted peak associated with relativistically-beamed matter in the approaching side of the accretion disk and an extended redshifted wing coming from matter very close to the black hole (Fig.~\ref{fig:reflectionlines}). Importantly, the X-ray reflection spectrum is truncated by the ISCO; within the ISCO the density of the accreting matter plummets as it accelerates radially inwards resulting in complete photoionization of ions in the plasma \citep{reynolds:08a}. Thus, the ISCO and hence black hole spin is imprinted on the X-ray reflection spectrum via the strength of the Doppler and gravitational broadening.

The development of the reflection picture, and the resulting techniques for measuring spin, has been observationally  driven from the start.  {\it EXOSAT} spectra of the first black hole X-ray binary Cygnus X-1 discovered a broadened iron line, prompting the suggestion and first models of Doppler/gravitationally-shifted iron fluorescence from the inner accretion disk \citep{fabian:89a}. The leap in spectral resolution possible with the {\it Advanced Satellite for Cosmology and Astrophysics (ASCA)} enabled the discovery of the broadened and skewed line expected from a relativistic accretion disk, first in the AGN MCG--6-30-15 \citep{Tanaka:95a} and then in many other bright AGN \citep{nandra:97a}. These early studies assumed iron line emission that truncates at the ISCO of a non-spinning black hole, but a detailed analysis of the low-flux state of MCG--6-30-15 revealed more extreme gravitational redshifting, necessitating either a rapidly spinning black hole \citep{iwasawa:96a,dabrowski:97a} or substantial line emission from within the ISCO \citep{reynolds:97a}. The case for a rapidly spinning black hole strengthened as it became clear that a significant fraction of the accretion energy is radiated in the X-ray corona thereby over-ionizing the region within the ISCO \citep{young:98a,reynolds:08a}. With the launches of {\it Chandra}, {\it XMM-Newton}, {\it Suzaku}, and more recently {\it NuSTAR}, the case for extreme gravitational redshifts and hence rapid black hole spin became stronger \citep{wilms:01a,fabian:02a,brenneman:06a,brenneman:11a,risaliti:13a} and was extended into the stellar-mass black hole realm \citep{miller:02a,miller:04a}.

\subsubsection{Modern methodology for X-ray reflection spin measurements} 

In parallel with these observational developments, and building upon pioneering work by \cite{basko:78a} and \cite{george:91a}, detailed models of the rest-frame X-ray reflection spectra from photoionized disk surfaces were developed \citep{ross:93a,ballantyne:01a,garcia:14a}. The modern methodology constructs an accurate model of the accretion disk reflection spectrum by convolving full rest-frame X-ray reflection spectra with the relativistic (Doppler/gravitational) broadening and redshifting effects. One needs to know the radial profile of coronal irradiation across the accretion disk surface in the construction of these models. Much of the early work (as well as some contemporary analyses) treated this empirically as a broken powerlaw form truncating at the ISCO.  But these empirical models often find highly centrally-concentrated irradiation profiles, with inner disk irradiation profiles that are $r^{-5}$ or steeper.  This suggests a ``lamppost'' geometry --- a compact source above the accretion disk and close to the black hole spin axis \citep{martocchia:96a,reynolds:97a,miniutti:04a}. Today, the strongest evidence for a lamppost corona comes the observed time delays between variations in the direct coronal emission and the X-ray reflection spectrum which are interpreted as light travel time delays between the disk and the elevated corona \citep{reynolds:99a,fabian:09a,kara:16b}.  It is increasingly common to adopt the lamppost model when fitting models to extract spin parameters. 

Overall, disk reflection models are characterized by the inclination of the accretion disk to the observer $i$, the ionization parameter of the inner disk $\xi$, the elemental abundance of the accretion disk (usually phrased in terms of the iron abundance $A_{\rm Fe}$), the height of the coronal source $h_X$ (in units of $r_g$), the spectral shape of the irradiating X-ray continuum (i.e. the photon index $\Gamma$ of the X-ray power-law), and the spin of the black hole $a$. Note that the mass of the black hole is not relevant to these spectral models since the velocity field of the accretion disk and the gravitational potential is scale-free once put in terms of gravitational radii $r_g$.

\subsubsection{The statistical robustness of X-ray reflection based spin measurements}

The majority of spin measurements published to date use the X-ray reflection technique (see Section~\ref{results}). Given this, it is appropriate to spend some time discussing the robustness of this technique, both to statistical degeneracies between model parameters and model uncertainties. 

The parameter space for these disk reflection models is at least six-dimensional $(i, \xi, A_{\rm Fe}, h_X, \Gamma, a)$ and the likelihood function can have a complex structure. The ionization state of the accretion disk, in particular, affects the disk spectrum in a complex and non-monotonic manner making it a problematic parameter when fitting spectra.  Especially if one focuses on the hard X-ray spectrum ($>2\keV$) in order to avoid model ambiguities surrounding the soft-excess and absorption, we often find statistically acceptable solutions at both low- and high-ionization parameters, but statistically poor solutions at intermediate ionization states.  The two acceptable solutions may have different spins, and this ridge-line in likelihood space can prevent some of the simpler fitting techniques (such as steepest decent $\chi^2$ fitting) from ever finding the global best-fit if the starting point is on the wrong side of the ridge-line. Practically, this issue has probably not affected many published measurements to date as each object is still analyzed individually and care is usually taken to manually explore parameter space as fully as possible, but it must be addressed when attempting to systematize future analysis of large datasets \citep[as demonstrated explicitly in the studies of simulated data of][]{bonson:16a}.	

\begin{figure}[t]
\includegraphics[width=4.8in]{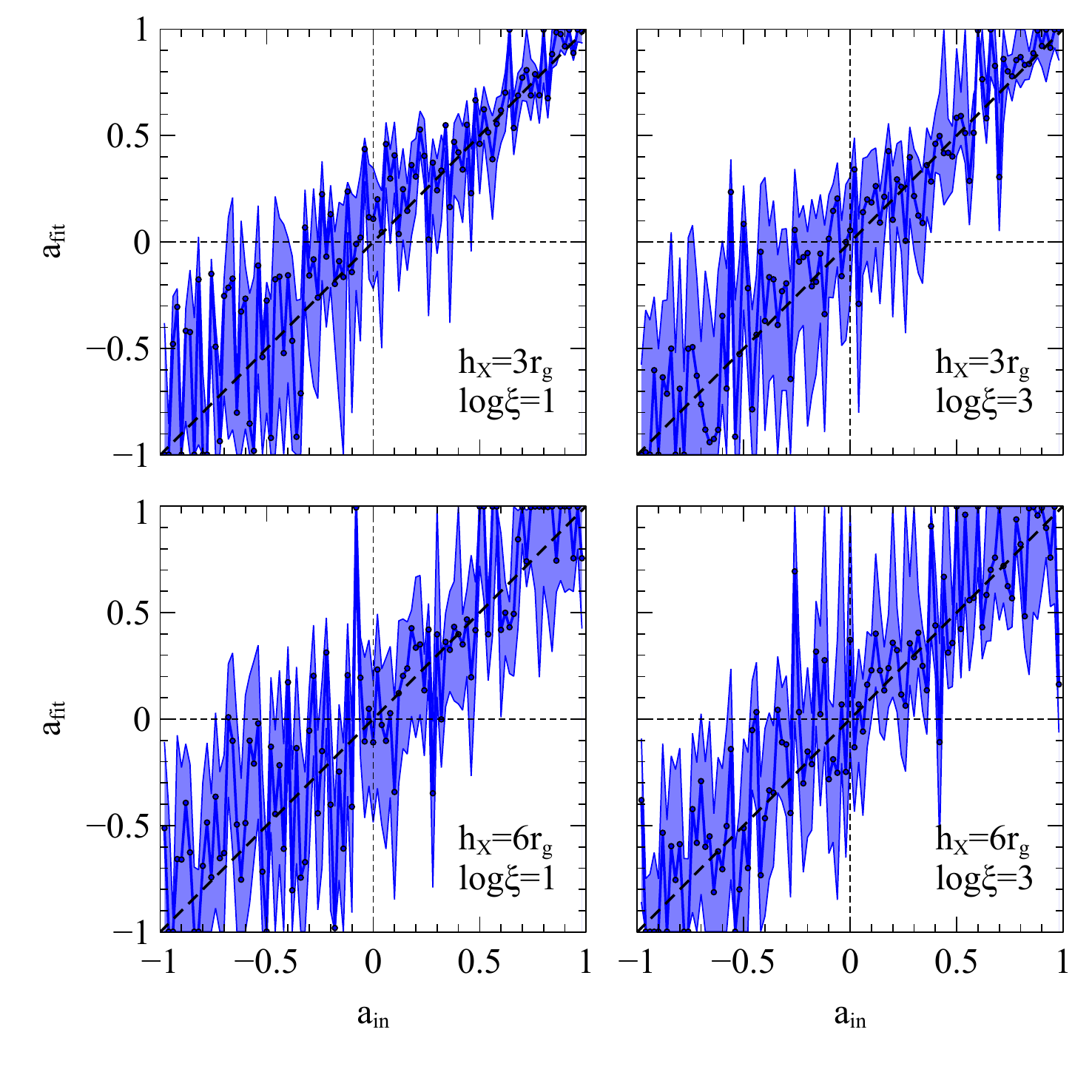}
\vspace{-0.5cm}
\caption{Example of the recoverability of spin signatures in simulated AGN X-ray reflection spectra.  The AGN's primary X-ray continuum is taken to be a powerlaw with photon index $\Gamma=1.9$ emitted from a point-like source on the black hole's spin axis at a height $h_X$ above the disk.  We assume that the accretion disk is viewed at an inclination of $i=30^\circ$, and that the irradiation of the disk by the X-ray source leads to a surface ionization parameter $\xi$ and an associated reflection spectrum.  We assume that the total observed 3-10\,keV flux of the AGN (primary$+$reflection) is $F_{3-10}=2\times 10^{-11}\ergpcmsqps$.  Then, for four pairs of $(h_X, \xi)$ and a dense grid of input black hole spins $a_{\rm in}\in(-1,1)$, we use the {\tt relxill} model \citep{garcia:14a} to produce a set of model AGN spectra that include the relativistically-broadened reflection spectra. For each model spectrum, we then simulate a 200\,ks {\it XMM-Newton} EPIC-pn observation and a 100\,ks {\it NuSTAR} observation.  We analyze the simulated spectra, using the 2--10\,keV band for {\it XMM-Newton} and 3--70\,keV band for {\it NuSTAR}.  These simulated spectra are fitted ``blind'' (i.e with arbitrary initial conditions) with the powerlaw$+$reflection model using standard  $\chi^2$-minimization techniques in order to recover the best fitting value and 90\% credible range for the spin, $a_{\rm fit}$. We see that retrograde spin can be difficult to constrain but prograde spin is generally recovered well, especially for low-down coronae ($h_X=3r_g$).  Central to the success of these fits is an iteration loop aimed at overcoming the ``ridge-line'' in likelihood space associated with intermediate values of the ionization parameter. }
\label{fig:spinrecovery1}
\end{figure}

Having recognized this issue, however, it is straightforward to build analysis threads that explore the full range of possible ionization parameters.  With that issue in hand, systematic explorations of simulated ``bare'' AGN (i.e. no intrinsic absorption) show that black hole spin can be faithfully recovered from high-quality single-epoch X-ray spectra provided that the coronal lamppost is sufficiently close to the disk \citep[see][as well as the author's study presented in Fig.~\ref{fig:spinrecovery1}]{kammoun:18a}. 

\subsubsection{Systematic uncertainties in X-ray reflection based spin measurements}

A more fundamental concern is the uncertainty in spectral models. There are two distinct aspects to this. First, there are approximations and assumptions made about the structure of the accretion disk when modelling the X-ray reflection spectrum. The radiative transfer calculations underlying the model rest-frame reflection spectrum treat the disk photosphere as a constant density slab, clearly an oversimplification although it remains unclear whether this makes a practical difference to the result. Also, the modeling of the relativistic Doppler/gravitational frequency shifts treat the disk as being entirely in the $\theta=\pi/2$ plane, i.e. flat with zero thickness and aligned with the black hole spin axis. While the alignment assumption is potentially justified due to the grinding action of Lens-Thirring torques \citep{bardeen:75a}, real disks will certainly have finite thickness (equation~\ref{eqn:h}). This changes the irradiation profile across the disk, the gravitational redshift of the disk photosphere, and potentially the moves the location of the final plunge away from the ISCO \citep{reynolds:08a,taylor:18a}.  Neglecting these effects will lead to a systematic error in the derived spin which, although modest as long as $h<r_g$, may exceed the statistical errors on spin for high-quality datasets \citep{taylor:18a}.

The second and more serious issue is whether we have correctly attributed the observed spectral structure to inner disk reflection. In many cases, the X-ray spectrum of an AGN is complex, including X-ray reflection/emission components from more distant/low-velocity matter as well as obvious line-of-sight absorption by outflowing photoionized gas \citep[generically referred to as warm absorbers; ][]{reynolds:97b,mckernan:07a}. Scattering and absorption by high-column densities of photoionized gas can mimic the signatures of accretion disk reflection, especially in medium resolution X-ray spectra that cut-off at 10\,keV such as can be obtained by the CCD spectrometers on {\it XMM-Newton}, {\it Chandra} and {\it Suzaku}. The degeneracy between absorption and disk reflection models is particularly pernicious if the absorbing photoionized matter only partially covers the primary X-ray source, e.g., if the central X-ray source is viewed through a ``mist'' of absorbing clouds, each of which is smaller than the size of the X-ray corona \citep{millerl:08a}. 

The result has been a lively debate about the relative role of complex absorption and relativistic disk reflection in AGN. Some objects clearly display both spectral signatures which, with high signal-to-noise data across a sufficiently broad bandpass, can be separately modelled and disentangled especially if the Compton reflection hump can be identified. A case in point is NGC~1365 which is well known to experience highly variable absorption as circumnuclear clouds pass in front of the X-ray source.  A campaign of simultaneous {\it XMM-Newton} and {\it NuSTAR} observations obtained a sequence of high-quality 0.5--70\,keV spectra which clearly revealed the changing absorption, but also the broad iron line and Compton reflection hump from an accretion disk around a rapidly spinning black hole \citep{risaliti:13a}. In an important check of the consistency of the spectral modelling, the inferred black hole spin and disk inclination remained invariant as the source experienced deep absorption events \citep{walton:14a}.  Independently, the observation of reverberation time delays between continuum fluctuations and the X-ray reflection spectrum \citep{fabian:09a}, including the broad iron line \citep{kara:16b}, has been a powerful tool for disentangling reflection and absorption models.   


\subsection{Spin from thermal continuum fitting}\label{continuum}

The thermal continuum fitting (CF) method is based on the fact that the spin of a black hole will influence the temperature of the inner disk, with prograde accretion onto a rapidly rotating black hole yielding the highest temperatures \citep{zhang:97a,mcclintock:14a}. This again follows from the fact that the spin influences the location of the ISCO, and hence the effective inner edge of the accretion disk. For higher spins, the smaller ISCO means that more binding energy is extracted from each gram of accreting matter, heating the inner disk to a higher temperature. Within the context of a specific accretion disk model, this can be quantified and turned into a precision tool for determining black hole spin.

\subsubsection{The physics of thermal disk emission and its dependence on black hole spin}

The fidelity of the spin measurements from CF is tied closely to the accuracy of the physical model for the accretion disk. For this reason, it most readily applies to systems with geometrically-thin, optically-thick, steady-state accretion disks with weak X-ray coronae. The structure of such accretion disks was calculated by \cite{novikov:73a} \citep[in a relativistic generalisation of the Newtonian model of][]{shakura:73a}.  The principal assumptions underlying the  \cite{novikov:73a} (hereafter NT) models are that (i) the system is in a steady state with an inwards mass flux that is constant with radius (i.e. mass loss due to disk winds is negligible), (ii) the accreting matter loses angular momentum via stresses internal to the disk (i.e. external torques due to a large scale magnetic field are negligible), and (iii) the energy dissipated in the flow is radiated locally. The last of these assumptions, the so-called radiatively-efficient condition, is justified provided that the accretion rate is in the range  $\dot{M}\sim 0.01-0.3\dot{M}_{\rm Edd}$ and naturally leads to a geometrically-thin and optically-thick disk structure.  This is the same Goldilocks-zone as employed for the X-ray reflection method, although reflection requires a strong X-ray corona whereas the most straightforward application of the CF-method assumes a weak corona.

Finally, we need to specify an inner boundary condition for the disk model, and we commonly assume that the internal stresses go to zero at the ISCO (the so-called zero-torque boundary condition). Given these assumptions, the radial profile of dissipated power per unit area of the disk $D(r)$ is completely determined by the mass, spin, and mass accretion rate, and is independent of any details of the complex MHD turbulent stresses that are thought to transport the angular momentum.  A closed form expression for $D(r)$ for the fully-relativistic theory can be written down \citep{novikov:73a} but is cumbersome.  Intuition is best gained from the Newtonian result,
\begin{equation}
D(r)=\frac{3GM\dot{M}}{4\pi r^3}\left( 1-\sqrt{\frac{r_{\rm isco}(a)}{r}}\right),
\end{equation}
\citep{shakura:73a,pringle:81a}.  Far from the ISCO ($r\gg r_{\rm isco}$), the disk locally dissipates/radiates three time more energy that is given up gravitationally by the local inflowing matter. The additional energy comes from the work done by viscous torques.  To compensate, the inner disk dissipates much less than the energy given up by the accreting matter; this energy is transported outwards in the disk by viscous forces.

The dissipated energy emerges from the disk photosphere as thermal radiation with effective temperature given by $D(r)=2\sigma_{\rm SB}T_{\rm eff}^4(r)$, where the factor of two accounts for the two radiating sides of the disk (i.e. the top and bottom of the disk).  Evaluating, again using the Newtonian model in order to gain intuition, we obtain
\begin{equation}\label{eqn:teff}
T_{\rm eff}=3.3\times 10^7\eta^{-1/4}\left(\frac{M}{10\Msun}\right)^{-1/4}\left(\frac{L}{L_{\rm Edd}}\right)^{1/4}\left(\frac{r_g}{r}\right)^{3/4}\left(1-\sqrt{\frac{r_{\rm isco}}{r}}\right)^{1/4}\,\K,
\end{equation}
where $L/L_{\rm Edd}=\dot{M}/\dot{M}_{\rm Edd}$ is the Eddington ratio of the system.  In our simple Newtonian disk model, the radiative efficiency will simply be $\eta=r_g/2r_{\rm isco}$.  
Maximizing equation~\ref{eqn:teff} over radii, we find that the effective temperature will reach a peak value, 
\begin{equation}
T_{\rm eff,max}=6.16\times 10^6\eta^{-1/4}\left(\frac{M}{10\Msun}\right)^{-1/4}\left(\frac{L}{L_{\rm Edd}}\right)^{1/4}\left(\frac{r_{\rm isco}}{r_g}\right)^{-3/4}\,\K,
\end{equation}
at a radius of $r_{\rm peak}={25\over 9}r_{\rm isco}$.  Using a fiducial Eddington ratio of 10\%, we see that the thermal spectrum of an accreting stellar-mass black hole ($M\sim 10\Msun$) peaks at $T_{\rm eff}\approx {\rm few}\times 10^6\K$, corresponding to $kT\approx {\rm few}\times 0.1\keV$ i.e. soft X-rays.  On the other hand, the thermal spectrum of an AGN ($M\sim 10^{6-9}\Msun$) will peak at $T_{\rm eff}\sim 10^{5-6}\K$ corresponding to $kT\sim 10-100\eV$, i.e. the extreme ultraviolet (EUV).  The fact that the EUV is very heavily extinguished by the interstellar medium of our Galaxy, immediately makes it challenging to study the peak of the thermal emission in many AGN. 

\begin{figure}
\hbox{
\hspace{-3.8cm}
\includegraphics[width=0.5\textwidth]{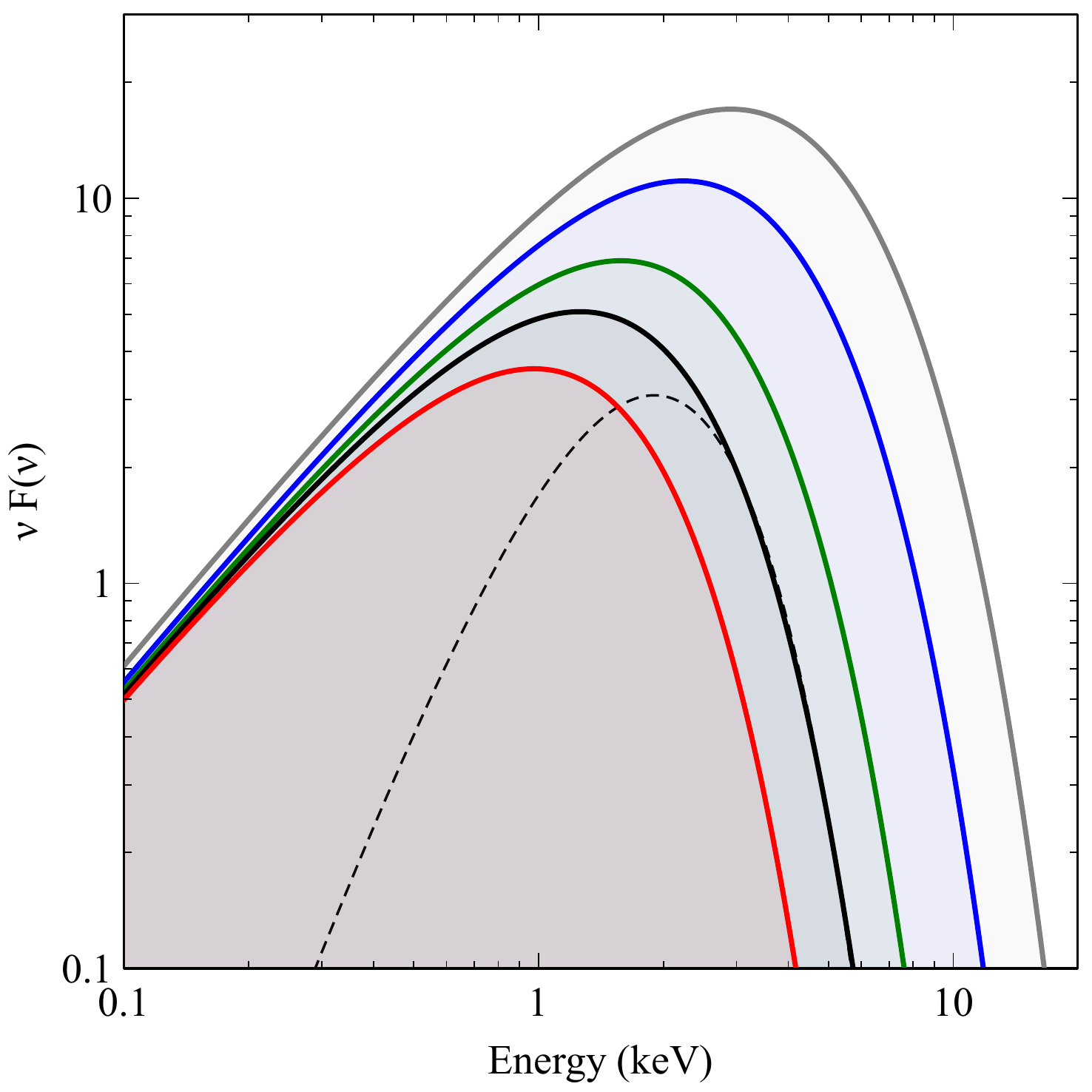}
\hspace{-6.5cm}
\includegraphics[width=0.5\textwidth]{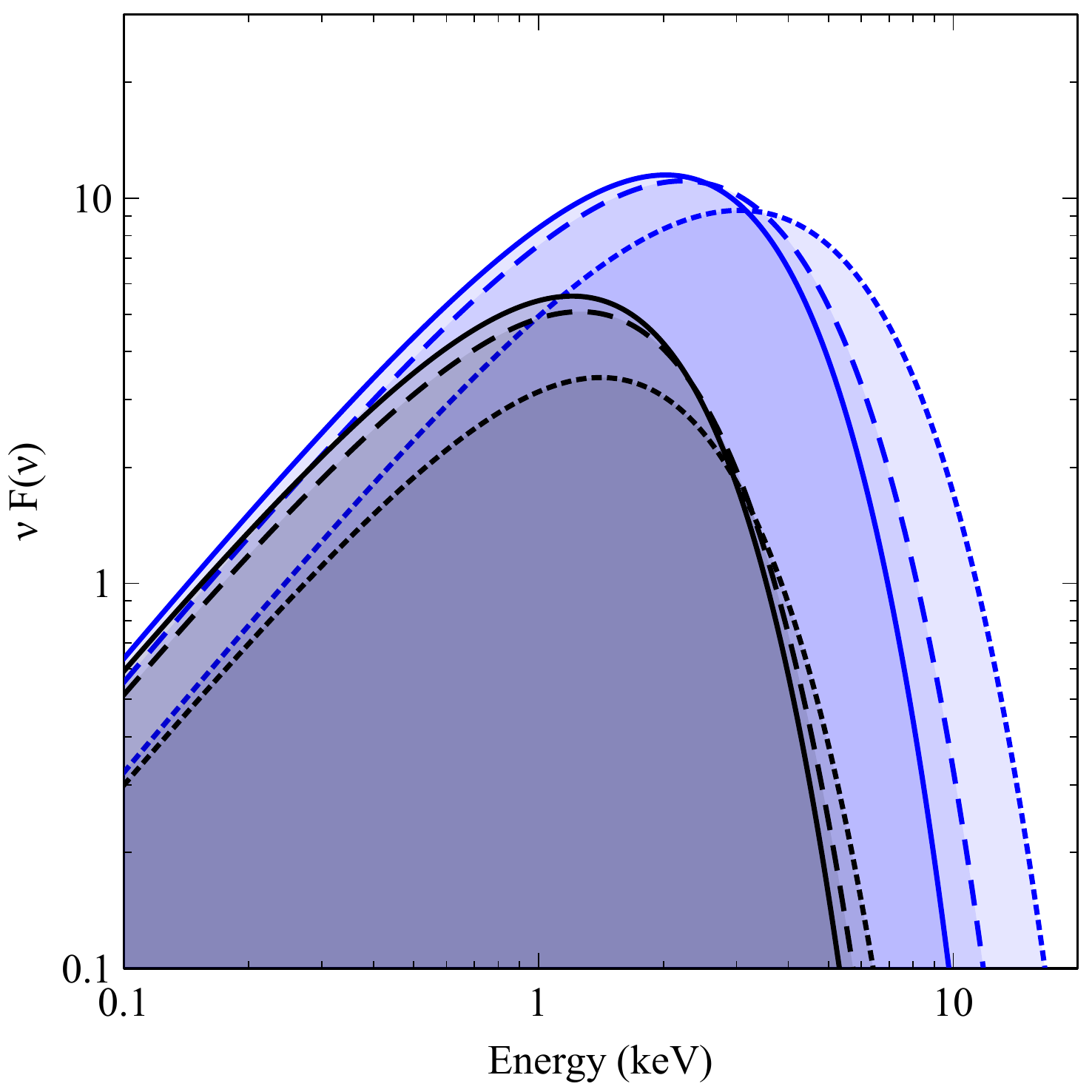}
}
\caption{Model thermal spectra for an accretion disk around a 10\Msun black hole, computed using the fully relativistic formalism of \cite{novikov:73a} as implemented in the {\tt kerrbb} spectral model \citep{li:05a} incorporated into the XSPEC spectral fitting package.  {\it Left panel : }Illustration of the spin dependence of thermal disk spectra. Shown here are models with $a=-0.9$ (red), $a=0$ (black), $a=0.5$ (green), $a=0.9$ (blue) and $a=0.998$ (grey).  All other parameters are held fixed across these models, with inclination $i=30^\circ$, $\dot{M}=2.6\times 10^{18}\gps$ (corresponding to 10\% of the Eddington limit for a Schwarzschild black hole with $\eta\approx 0.06$) and $f_{\rm col}=1.7$.  The dotted line shows a simple black body with $T=5.63\times 10^6\K$ (0.485\keV) and shows that the high-frequency tail of the full-disk thermal spectra are well approximated by a black hole of a single temperature. {\it Right panel : }Illustration of the effects of inclination on disk spectra.  Shown here are spectra with $i=0$ (solid), $i=30^\circ$ (dashed) and $i=50^\circ$ (dotted) for two spins, $a=0$ (black) and $a=0.998$ (blue).}
\label{fig:thermal_fitting}
\end{figure}

It can be seen from eqn~\ref{eqn:teff} that, for a given mass and Eddington ratio, $T_{\rm eff,max}$ will increase as the spin $a$ increases due to the corresponding decrease in $r_{\rm isco}$. Figure~\ref{fig:thermal_fitting} (left panel) shows fully-relativistic models of thermal disk spectra for parameters relevant to an accreting stellar-mass black hole in an X-ray binary system.  These disk-integrated spectra are much broader than a single temperature blackbody, but the peak and the high-frequency decline of the spectrum will be dominated by the hottest part of the disk. For a constant accretion rate, the spin dependence of $T_{\rm max}$ is strong, leading to changes by factors of $\sim 3-4$ in both the energy and normalization of the peak of the spectrum as one scans through the full range of spins $a\in (-1,1)$. 

An important technicality is that the emergent radiation from the accretion disk photosphere will never strictly be a blackbody due to scattering and absorption processes within the disk atmosphere. In the hot disks around stellar-mass black holes, energy-independent electron-scattering dominates and the local thermal spectrum will have an intensity of the form $I_\nu=B_\nu(f_{\rm col}T_{\rm eff})/f_{\rm col}^4$, where $B_\nu(T)$ is the usual Planck function describing black body radiation and $f_{\rm col}\sim 1-2$ is known as the color correction factor.  In other words, the local thermal spectrum has the shape of a blackbody with temperature $T_{\rm col}=f_{\rm col}T_{\rm eff}$ but has a total flux corresponding to a blackbody with the temperature $T_{\rm eff}$. Thus the peak of the observed thermal spectrum from the disk is shifted upwards by a factor of $f_{\rm col}$ compared to the n\"aive expectations, and it is obviously important to account for this fact in any attempts to measure black hole spin. Detailed radiative transfer calculations have enabled  estimates of $f_{\rm col}$ as a function of $T_{\rm eff}$ \citep[e.g.][]{davis:05a,davis:06a}.  Although remaining uncertainties in the structure of the disk atmosphere and hence $f_{\rm col}$ may introduce modest systematic errors on spin \citep{salvesen:20a}, this issue is largely in hand for black hole X-ray binaries. \begin{marginnote}[]
\entry{$f_{\rm col}$}{Color correction factor, approximately describing how a thermal/blackbody spectrum is distorted by scattering and absorption in the disk atmosphere.}
\end{marginnote}

The atmospheres of AGN disks are lower temperature, resulting in a much greater importance of atomic (highly frequency-dependent) scattering and absorption processes \citep{hubeny:01a}. While the spectral distortion is sometimes approximated using the color-correction factor formalism, the emergent spectrum is much more complex and structured. Together with the fact that the peak of the AGN thermal spectrum is in the hard-to-observe UV/EUV, the use of the thermal continuum fitting as a tool for measuring spins is much less developed for AGN than for accreting stellar-mass black holes.  

\subsubsection{Practical considerations for spin measurements using the thermal continuum-fitting method}

The successful application of the CF-method for measuring spin requires several ingredients. Firstly, we need the  system to be well-described by the NT-model, with no (or weak) winds, no (or weak) corona, and a spectrum that is dominated by thermal emission from the optically-thick accretion disk.  Spins are then derived using models with color correction factors $f_{\rm col}$ that are functions of temperature as derived from detailed radiative transfer calculations \citep{davis:05a,davis:06a,shafee:06a,mcclintock:14a}.  Black hole X-ray binaries cycle through distinct spectral states as they go through outburst cycles \citep{remillard:06a} and one of these states, the so-called thermal-dominant state, does indeed appear close to a pure thermal NT disk.  So, for stellar mass black holes, we can patiently monitor and wait for the system to enter the thermal dominant state and then use the resulting spectrum to determine spin.  Any equivalent state changes in AGN will occur on timescales of millennia, and so we must limit CF-studies to those AGN that instantaneously appear to be in a thermal dominant state.

Secondly, as is apparent from eqn.~\ref{eqn:teff}, the disk temperature depends on the accretion rate. This can be measured from the thermal spectrum via the normalization of the low-energy portion of the thermal spectrum, provided that the observed spectrum can be converted into luminosity units.  This, in turn, requires us to know the distance of the system from us.  This is rarely a problem for AGN which are almost always far enough to apply Hubble's Law to the measured cosmological redshift.  The distance to many black hole X-ray binaries, on the other hand, is often poorly known and inferred via indirect arguments (e.g. matching observed absorption column densities to models of Galactic gas distributions).  Increasingly, however, astrometric parallax measurements of the distance to  black hole X-ray binaries are becoming possible via radio interferometry \citep{reid:11a,atri:20a} and {\it Gaia} \citep{gandhi:19a}. 

Thirdly, the observed thermal spectrum depends on the black hole mass (as is explicit in eqn.~\ref{eqn:teff}) and the inclination of the accretion disk (Figure~\ref{fig:thermal_fitting}, right panel).  Inclination matters since we must characterize the Doppler shifts of the inner disk spectrum due to orbital motions.  These parameters must be determined by other means before spin information can be extracted from the thermal spectrum. In the case of black hole X-ray binaries, robust estimates of black hole mass and the inclination of the orbital plane can be obtained from optical measurements of the radial velocity and ellipticity-induced photometric variations of the companion star \citep{orosz:02a,orosz:07a}.  The resulting mass estimates typically have 10--20\% precision and, given the weak mass dependence of the disk temperature, rarely dominate the error budget.  The orbital inclination measurements are also typically precise (within a few degrees), but in the absence of other information we must make the further assumption that the black hole spin axis and the inner accretion disk are both aligned with the orbital angular momentum.  

Obtaining precision black hole masses and disk inclinations for AGN is more problematic.  For these actively accreting SMBHs, the most direct and robust method for determining mass comes from reverberation mapping of the broad line region \citep[BLR; ][]{blandford:82b,peterson:04a,bentz:06a}.  Given the unknown structure of the BLR, these masses are only considered accurate to within a factor of 2--3, and the uncertainties increase for systems close to the Eddington limit as radiation forces, rather than gravitation, start to dominate the BLR dynamics. \begin{marginnote}[]
\entry{BLR}{Broad line region, a region close to an accreting black hole producing prominent infrared, optical, and ultraviolet emission lines.}
\end{marginnote}The inclination of the inner disks in AGN is readily constrained if the system displays inner disk reflection signatures (Section~\ref{reflection}), or relativistic radio jets \citep{wills:95a}.  Unfortunately, the presence of either of these phenomena invalidates, at least at some level, the NT assumption of a dominant thermal disk that underlies this particular spin diagnostic.  Thus, for AGN that are strictly in the thermal-dominant state, the inner disk inclination is unknown.  We note that \cite{middleton:16a} \citep[also see][]{kinney:00a} has found a poor correlation between reflection-based disk inclinations and that of the large scale galactic disk

For the reasons outlined above, the CF-method for spin determination has been applied much more widely to black hole X-ray binaries than AGN.  The black hole X-ray binary results are discussed in Section~\ref{results}.  Here we just note that there have been a small number of studies looking at the CF-derived spins for SMBHs.  \cite{done:13a} identified PG1244$+$026 as an AGN in a thermal-dominant state and, using the fact that the thermal disk spectrum does not extend into the soft X-ray, set an upper limit of $a<0.86$ on spin.  \cite{capellupo:17a} examine the Seyfert galaxy NGC~3783 and the luminous quasar H1821$+$643 finding in both cases that CF implies $a>0.1$, consistent with the more constraining X-ray reflection measurements.  Using a time-dependent disk model, \cite{mummery:20a} examine the tidal disruption event ASASSN-15lh and find that, if the spectrum is thermally-dominated and the disrupted star is solar-like, the data requires a massive black hole $M\sim 10^9\Msun$ which is close to maximally spinning.  \begin{marginnote}[]
\entry{Tidal disruption event}{Destruction and subsequent accretion of a star that passes close to a supermassive black hole.}
\end{marginnote}

\subsubsection{Departures from the Novikov-Thorne disk model and systematic uncertainties on CF spin measurements}

The NT disk model is theoretically well-motivated, but does rest upon assumptions and approximations.  It is important to assess the systematic errors in derived spins that may result from departures from these assumptions and approximations.  

A key assumption of the NT model is that the disk locally radiates the energy dissipated in the accreting gas; in particular, any radial advection of internal energy is neglected. General viscous disk models show that the ratio of advected-to-dissipated energy is $\sim (h/r)^2$, and so we expect that the neglect of advection will become progressively more problematic as one considers higher accretion rates (hence thicker disks). Observations nicely agree with these theoretical expectations. \cite{mcclintock:06a} examined the CF-inferred spin of the Galactic microquasar GRS1915$+$105 as a function of its instantaneous luminosity.  Consistent values of spin ($a>0.98$) were recovered for all periods with $L/L_{\rm Edd}<0.3$.  For higher luminosity, however, the inferred spin monotonically decreased with luminosity qualitatively consistent with advective cooling of the inner disk.  \cite{steiner:10a} found very similar behaviour in the black hole X-ray binary LMC~X-1, showing that the effective inner radius of the thermal disk is constant at low-luminosity but increases with luminosity above $L/L_{\rm Edd}\approx 0.3$ exactly as expected from ``slim disk'' models that include advection \citep{straub:11a}.  

Even for thin ($L/L_{\rm Edd}<0.3$) disks, accreting matter will continue to emit even while undergoing its final plunge within the ISCO due to the advection of internal energy over the ISCO. This component is not included in the standard CF-analysis. \cite{zhu:12a} examine the thermal emission predicted from a set of General Relativistic MHD (GRMHD) simulations of thin accretion disks and find that the effective temperature profile remains flat such that $T\approx T_{\rm eff,max}$ within $r<r_{\rm peak}$, including inside of the ISCO.  However, much of the flow within the ISCO has a substantial inward radial velocity that relativistically beams much of its thermal emission into the black hole.  Together with the strong gravitational redshift of this region, emission from within the ISCO makes only a modest contribution to the observed spectrum and introduces systematic errors in spin measurements of $\Delta a<0.05$. We note that a tentative detection of thermal emission from within the ISCO has recently been reported in the black hole X-ray binary MAXI~J1820+070 by \cite{fabian:20a}.\begin{marginnote}[]
\entry{GRMHD}{General Relativistic Magnetohydrodynamics, description of the dynamics of a magnetized fluid with relativistic velocities in curved spacetimes.}
\end{marginnote}

Finally, contrary to the assumptions of NT, magnetic fields can tie plunging matter within the ISCO to the rest of the disk, leading to a violation of the assumption of zero-torque at the ISCO \citep{gammie:99a,reynolds:01a}.  The torque exerted from the plunging flow does work on the inner disk that is viscously transported outwards, leading to enhanced dissipation and an altered temperature profile across an extended part of the inner disk \citep{agol:00a}.  However, using $\alpha$-models that include ISCO torques, \cite{shafee:08a} have shown that the systematic error introduced by the uncertain presence of such torques is $\Delta a\approx 0.1$ for $a=0$, and significantly less for more rapid spins.  

\subsection{Other methods for obtaining spin of accreting black holes}\label{other}

So far we have focused on the X-ray reflection and CF methods for inferring black hole spin as they dominate the literature and have been subject to the most scrutiny.  A major limitation of both methods is that they rely upon the presence of a geometically-thin, optically-thick disk and hence can only be applied with confidence to systems in the Goldilocks-zone of accretion rates ($L/L_{\rm Edd}\sim 0.01-0.3$).  However, there are many other ways that spin can manifest in observations of accreting black holes, leading to a number of other proposed techniques for measuring spin to which we now turn our attention.  Due to limited space, our discussion here is necessarily brief but we refer to key review articles where possible.

\subsubsection{Spin from quasi-periodic oscillations}\label{qpos}

Many black hole X-ray binaries and a few AGN  show quasi-periodic oscillations (QPOs), i.e. broadened peaks in the temporal power-spectrum of their X-ray emission. The temporal frequencies of QPOs must be related to specific frequencies of the underlying accretion flow and black hole spacetime, and thus must encode spin information. The phenomenon is much better characterized in black hole X-ray binaries, so we limit our discussion here to those systems.  \begin{marginnote}[]
\entry{QPO}{Quasi-periodic oscillation, variability that is approximately but not exactly periodic in nature.}
\end{marginnote}
The nature and properties of QPOs in X-ray binaries has a rich phenemenology, being strongly dependent upon the spectral state \citep[see][for review]{ingram:19a}. Broadly, there are three types of low-frequency QPOs (LFQPOs; frequency $\nu\sim 0.1-10\Hz$), catagorized as type-A/B/C depending upon their strength, frequency shifts, frequency width, and relationship with the underlying broad-band noise.  High-frequency QPOs (HFQPOs; $\nu\sim 100\Hz$) are rarer, being occasionally seen at high accretion rates.  HFQPOs have approximately fixed frequency, leading to the hypothesis that HFQPOs are tied to fundamental frequencies of the underlying spacetime.  

To extract spin measurements from QPO observations requires a theoretical model. Using the {\it Rossi X-ray Timing Explorer} (RXTE), \cite{strohmayer:01a} found two HFQPOs in the source GRO~J1655--40 with frequencies $\nu=300, 450\Hz$ and, assuming that the upper frequency is bounded by the ISCO orbital frequency, deduced $a>0.15$.  Noting the 2:3 ratio of these frequencies, \cite{abramowicz:01a} suggested they are manifestations of a parametric resonance between the orbital and radial epicyclic frequencies which, for a given measured mass, picks out a  preferred radius and spin parameter.  Applying this to GRO~J1655--40, \cite{abramowicz:01a} inferred a spin $a=0.2-0.67$.  Other suggestions include HFQPOs as global $g$-modes of the inner disk \citep{nowak:97a}, local pressure-driven oscillation modes \citep{reynolds:09a}, and rotational modulation of the base of a Blandford-Znajek jet \citep{mckinney:12a}.  Each of these suggestions provide a distinct mapping of HFQPO frequency to spin.

More recent attention has focused on LFQPOs.  Motivated originally by the QPOs seen in neutron star X-ray binaries, the relativistic precession model \citep[RPM; ][]{stella:99a} identifies the (type-C) LFQPO with the test-particle Lens-Thirring precession frequency of some characteristic radius, and the lower/upper HFQPO frequency with the test-particle periastron precession and orbital frequencies of that radius. Applying the RPM to GRO~J1655--40, \cite{motta:14b} derived a black hole mass ($M=5.31\pm 0.07\Msun$) in agreement with optical measurements, but a spin ($a=0.290\pm 0.003$) that was significantly smaller than inferred from CF \cite[$a=0.65-0.75$;][]{shafee:06a} and X-ray reflection \cite[$a>0.9$;][]{reis:09a}.  

The major limitations of the RPM are that (i) it ascribes special significance to some arbitrary  radius (which becomes a parameter of the model) and (ii) it is based on pure test-particle dynamics thereby neglecting any non-gravitational physics.  By constructing some of the first GRMHD simulations of black hole accretion, \cite{fragile:07a} found that a geometrically-thick accretion disk whose angular momentum is mis-aligned to that of the Kerr black hole will undergo Lens-Thirring induced coherent precession. The Lens-Thirring torques and resulting test-particle precession frequencies are strongly decreasing functions of radius, but the (wave-like) transmission of the warp angular momentum through the disk lead to a coherently precessing structure with a frequency that is a weighted average of the test-particle frequencies.  This is now the leading model for the type-C LFQPOs seen in the low-hard state of black hole X-ray binaries \citep{ingram:09a}; in such a state, the inner disk is believed to be a radiatively-inefficient, optically-thin, geometrically-thick structure that can undergo coherent Lens-Thirring induced precession.  A prediction of this model is the existence of a moderately broad iron line in the X-ray spectrum that ``rocks'' with the QPO-frequency due to the non-axisymmetric X-ray illumination of the outer radiatively-efficient, optically-thick, geometrically-thin disk by the precessing inner structure \citep{ingram:16a}.  Indeed, such iron line behaviour was noted in RXTE data of GRS1915+105 and interpreted as a Lens-Thirring induced warped disk prior to the theoretical underpinnings given by GRMHD simulations \citep{miller:05a}. While more theoretical and observational work is required to understand the transition radius and coupling of the inner/thick and outer/thin disks, LFQPOs hold promise as a future precision probe for spin.

\subsubsection{Radiative efficiency}\label{efficiency}

For a steady-state accretion flow, the radiative efficiency is defined as 
\begin{equation}\label{eqn:eta}
\eta=\frac{L}{\dot{M}c^2}
\end{equation} 
where $L$ is the total (i.e. frequency- and solid-angle integrated) radiated luminosity as measured by a distant observer.  For a NT-like disk, the efficiency is simply given by $\eta=1-e_{\rm isco}/c^2$ where $e_{\rm isco}$ is the total energy of a unit mass in circular orbit at the ISCO.  Thus, under these conditions, $\eta$ is purely a function of spin ranging from $\eta\approx 0.057$ for a non-spinning black hole to $\eta\approx 0.42$ for $a=1$. 

Given independent measurements of the mass accretion rate and luminosity, eqn.~\ref{eqn:eta} trivially gives the efficiency and hence the spin.  \cite{davis:11a} point out that the optical flux from an AGN is dominated by the regions of the accretion disk at $r\sim 100r_g$ and so is good measure of the mass accretion rate, being unaffected by spin which only manifests in signals from the innermost disk.  They proceed to examine a sample of 80 quasars, finding that efficiency increases as a function of mass from $\eta\sim 0.03$ for $M\sim 10^7\Msun$ to $\eta\sim 0.4$ for $M\sim 10^9\Msun$.  Taken at face value, this would suggest a transition from non-rotating or even counter-rotating black holes at the low-mass end to rapidly spinning black holes at the highest masses.  This conflicts with AGN population trends seen in both X-ray reflection (Section~\ref{results}) and jet-power (Section~\ref{jet}) studies. One possibility is that the NT-model is a poor description of AGN disks, an assertion that has observational support from studies of quasar microlensing (Section~\ref{microlensing}).  In particular, UV-line driven winds may carry away significant amounts of matter thereby altering the disk structure \cite{laor:14a}.   In addition, the bolometric luminosity of such systems is often dominated by the unobservable EUV wavelengths, requiring uncertain extrapolations that may introduce biases as a function of mass.  

Efficiency arguments can also be applied in an integral sense.  Integrating eqn.~\ref{eqn:eta} over cosmic time gives
\begin{equation}
\langle\eta\rangle=\frac{(1+\langle z\rangle) \epsilon_{\rm bol}}{\rho_{\rm bh}},
\end{equation}
where $\epsilon_{\rm bol}$ is the average energy density of quasar light in the Universe today, $\rho_{\rm bh}$ is the average space density of black hole mass today, and  $\langle z\rangle$ the average redshift of the AGN emission (which is relevant since radiation and mass dilute differently as the Universe expands).  \cite{soltan:82a} used the observed optical quasar luminosity function and X-ray background to estimate $\epsilon_{\rm bol}$, showing that relatively efficient accretion $\langle\eta\rangle\sim 0.1$ was required to give reasonable black hole mass densities.  Equipped with updated average quasar spectra and new estimates of black hole masses, \cite{elvis:02a} found $\langle\eta\rangle>0.15$,  suggesting that the bulk of the SMBH population had to be rapidly spinning.  This result \cite[recently updated by][who has highlighted the importance of selection biases when extrapolating measured black hole masses to the full population]{shankar:20a} remains an important and robust integral constraint on the spin of the average SMBH population over cosmic time.

\subsubsection{Jet power}\label{jet}

The current prevailing paradigm is that the relativistic jets displayed by some AGN and black hole X-ray binaries are powered by the magnetic extraction of the black hole's spin energy.   \cite{blandford:77a} provide the theoretical foundations for this model showing that, at least for modest spins, the power extracted from a spinning black hole whose horizon is threaded by a magnetic field $B_H$ is
\begin{equation}\label{eqn:bz}
L_{\rm jet}\approx\frac{1}{128}B_H^2r_{\rm evt}^2a^2c,
\end{equation}
where we have made an ``impedance matching'' assumption that maximizes jet power by requiring that the horizon threading field rotates with half the angular velocity of the horizon \citep[see discussion in][]{blandford:77a}.  The analytic treatment of \cite{blandford:77a} is formally only valid for small spins since it is a perturbation analysis around a non-spinning solution, but GRMHD simulations show that $L_{\rm jet}\propto a^2$ remains valid up to quite high spins \citep[although higher order terms may contribute at the highest spins;][]{tchekhovskoy:10a}.  Observational support for the spin-driven jet paradigm comes from \cite{narayan:12a} who find a correlation between the peak radio luminosity (taken as a measure of peak jet power) and the CF-inferred black hole spins.   However, we note the dissenting viewpoint of \cite{russel:13a} who point out that different radio outbursts of a given source can peak at very different fluxes and hence challenge the existence of a radio-spin correlation once all available radio/spin data are used.  We conclude that it is still too early to based quantitative measures of spin on radio observations of individual black hole X-ray binaries.

The most systematic attempts to measure black hole spin via jet-power have been in the realm of AGN \citep{daly:09b}.   To invert eqn.~\ref{eqn:bz} in order to determine spin requires a knowledge of the black hole mass, jet power and $B_H$. Black hole mass is commonly estimated by applying known correlations between SMBH mass and galaxy properties (stellar mass, luminosity, or velocity dispersion) or virial estimates based on observations of the BLR.  Jet power can be challenging to measure, with the most robust values coming from estimates of the ``p\,dV'' work done as the jets excavate a cavity in the surrounding medium \cite{mcnamara:09a}.  The horizon-threading magnetic field $B_H$ is, however, unconstrained by observation and so one must remove this degree of freedom either with a simple assumption (e.g. $B_H\propto a$) or by relying upon a model of the inner accretion flow \citep{daly:11a}.  The power of this technique is that it can be applied to large samples of jetted-AGN \citep{daly:11a,daly:19a}, many of which exhibit accretion states that make the X-ray reflection or CF methods inapplicable.  The spins inferred from this technique generally respect but fill up to the $|a|<1$ bound; this is an important consistency check that need not have been true if any of the assumptions underlying the method were flagrantly violated. Given that we stand on the threshold of next generation radio surveys (e.g. LOFAR, SKA), this may prove to be an important technique for future extragalactic spin measurements.  

\subsubsection{Quasar microlensing}\label{microlensing}

A fortuitous alignment of a massive galaxy between the Earth and a distant ($z\sim 1-2$) AGN can result in strong gravitational lensing and the appearance of multiple images of the AGN. For typical parameters, there are either two or four images of the AGN separated by $\sim 1$\,arcsec and they are each viewed through the stellar component of the intervening/lensing galaxy. Each individual AGN image is then subject to microlensing by the gravitational fields of the stars in the lensing galaxy leading to independent brightness fluctuations that will be imposed on top of any intrinsic variability associated with the AGN itself.  The typical magnitude of the microlensing-induced variability depends on the size of the emitting region in the AGN relative to the typical size-scale of the lensing caustic network with larger the emitting regions leading to reduced microlensing modulation.

Putting together high-cadence multi-wavelength monitoring of strongly-lensed AGN with careful modeling of microlensing, we can map the frequency-dependent surface-brightness profile of the accretion disk \citep{agol:99a}.  Interestingly, this approach has shown that the optical size of many AGN disk are $2-3\times$ greater than suggested by eqn.~\ref{eqn:teff} \citep{morgan:10a}, a result that currently stands as one of the most significant challenges to standard accretion disk theory in AGN.  The sub-arcsecond imaging capabilities of {\it Chandra} allow the separation of the individual lensed images and hence microlensing studies in the X-ray band \citep{morgan:08a}, providing completely independent confirmation that the X-rays originate from a compact ($r<10r_g$) source.  

The possibility of using these signals to infer black hole spin comes from the fact that X-ray microlensing can, for a while, preferentially magnify the X-ray reflecting inner accretion disks.  Current methodologies attempt to constrain spin by characterizing either the energy distribution of transient iron line features \citep{chartas:17a} or the enhancement in the iron line equivalent width compared with non-lensed AGN \citep{dai:19a}.  Current results are limited by the effective area and count rates of {\it Chandra}, but these techniques hold great promise when the next-generation of large-area, sub-arcsecond X-ray facilities are deployed on the large number of lensed quasars expected to be discovered by the Vera C. Rubin Observatory.  

\subsubsection{Spin from direct imaging}\label{imaging}

The notion of using mm-band Very Long Baseline Interferometry (VLBI) to image horizon-scale emission around nearby SMBHs has been around for some time \citep{falcke:00a}.  This became a reality in just the past two years with the publication of the Event Horizon Telescope (EHT) images of horizon scale structures around the SMBH at the center of the giant elliptical galaxy M87 \citep{eht:19a}.  The image is dominated by a ring that corresponds to the photon circular orbit (see Section~\ref{physics}). \begin{marginnote}[]
\entry{EHT}{Event Horizon Telescope, a global interferometric network of radio telescopes capable of imaging horizon scale structure in a smaller number of nearby supermassive black holes.}
\end{marginnote}The ring is clearly asymmetric, and comparison with libraries of GRMHD simulations suggest that, provided the black hole has any appreciable spin at all, the asymmetry is due to relativistic beaming of plasma that is corotating with the black hole within the ergosphere.  Given the other parameters at play in these models such as the unknown inclination, magnetic fluxes, and electron temperatures, the EHT imaging data alone do not constrain the spin strongly, with adequate agreement between models and data possible for $-0.94<a<+0.94$ \citep{eht:19b}.  Including the fact that M87 possesses a powerful jet leads to a preference for models with high spin \citep{eht:19b}, although, of course, this is a theoretical prejudice and not a measurement, and they further note that {\it none} of their GRMHD models produce a jet with the power of the time-averaged M87 jet ($L_{\rm jet}\sim 10^{44}\,{\rm erg}\,{\rm s}^{-1}$).  Still, future high-fidelity interferometric imaging of some special targets (Sgr~A$^*$ and M87), especially if baselines can be extended via space-based antennae, clearly holds great promise in probing the spin.

\section{BLACK HOLE SPIN FROM GRAVITATIONAL WAVE SIGNALS}\label{gwa}

Thus far, we have discussed measurements of black hole spin from accreting black holes using EM observables. GWs provide a fundamentally different window on the gravitational physics of black holes. GW signatures of relativistic gravity, including spin, are ``clean'' in the sense that they are not subject to the complexities affecting our understanding of accretion flows. On the other hand, the imprints of spin on GWs can be subtle and, at the current level of sensitivity provided by the Laser Interferometer Gravitational Wave Observatory (LIGO), there are still only a small number of strong spin constraints. This section will focus on the current results from the LIGO-Virgo Consortium (LVC).  An immediate problem that arises in reviewing this area is that the discussion is out of date almost immediately --- the discussion presented here captures the state of the field as of early-September 2020.  We will look ahead to the future prospects of GW astronomy, especially those enabled space-based GW observatories, in Section~\ref{future}.

In the accreting systems discussed in Section~\ref{accreting}, the black holes are expected to be well described by the Kerr metric, with the accretion disk having little impact on the gravitational physics close to the black hole. By contrast, today's GW  studies focus on black holes that are in binary systems, either black-hole/black-hole or black-hole/neutron-star binaries.  GWs carry away energy and angular momentum from such binaries, leading to a gradual decay of the orbit and eventually a collision/merger of the two components.  On 14th September 2015, LIGO detected GWs from the merger of a 36\Msun\ black hole with a 31\Msun\ black hole to form a final (spinning) 63\Msun\ black hole \citep[with the remaining mass-energy being carried away by GWs;][]{abbott:16a}.  In addition to being the first robust direct detection of GWs, this source (designated GW150914) was the first ever observation of a BBH.  The masses of the two black holes were both significantly higher than those found in black hole X-ray binaries, highlighting the need to consider a range of formation scenarios beyond simple binary star evolution (Section~\ref{results}).  The need to explore the diversity of formation scenarios has further increased the relevance and interest in black hole spin.

At the time of writing (September 2020), LIGO has completed three observing runs (O1--3), the last two of which were joined by the Advanced Virgo detector.  The Kamioka Gravitational Wave Detector (KAGRA), the first GW to be built underground and to use cryogenic mirrors, came online and joined LIGO-Virgo O3 in late-February 2020, one month before operations were suspended due to the Covid-19 pandemic.  In the first and second observing runs, O1/O2, a total of 10 merging BBHs were discovered \citep{abbott:19a}.  All of these binaries possessed components that had approximately equal mass, a selection effect that will be discussed further below and is relevant to the derived spin constraints.  The full catalogue of results from O3 has yet to be published, but the LVC have published an analysis of three particularly interesting O3 events, two of which have yielded the strongest GW-based constraints on black hole spin to date and so will be discussed individually below (Sections~\ref{gw190412}-\ref{gw190814}; the third O3 event is very likely a binary neutron star merger and hence lies outside the scope of our discussion). 

In the rest of this Section, we describe the three stages in the life of a merging BBH. We then discuss how the spins of the two pre-merger black holes, as well as that of the final remnant black hole, are imprinted on the GW signal.  We proceed to summarize our current knowledge of both the progenitor and final black hole spins, highlighting the important role of the assumptions (i.e. priors) used in when drawing inferences from the GW data.  Even with a handful of results, it now seems clear that the population of black holes detected by GW astronomy are very different to those that inhabit X-ray binary systems..\begin{marginnote}[]
\entry{Priors}{Initial assumption about the likely values of some quantity}
\end{marginnote}

\subsection{The three stages of a merging binary black hole}\label{stages}

To state the obvious, there are three distinct stages to a merger of a BBH; before, during, and after merger.  Here, we describe the basic dynamics of each stage, setting the scene for the discussion of spin constraints in Sections~\ref{inspiral} and \ref{merged}.  

\subsubsection{Inspiral (before)} In the inspiral stage, the two black holes are well separated and orbit their common center of mass.  If the two black holes have masses $M_1$ and $M_2$ and are separate by a distance $r\gg r_g$ (so that orbits are approximately Newtonian and GW emission is dominated by the quadrupole term), the rate of loss of energy to GWs is given by
\begin{eqnarray}\label{eqn:gw1}
L_{\rm GW}&=&\frac{32G^4}{5c^5}\frac{M_1^2M_2^2(M_1+M_2)}{r^5}\\
&\approx &1.5\times 10^{49}\left(\frac{\nu}{0.25}\right)^2\left(\frac{r}{100r_{gT}}\right)^{-5}\,\ergps
\end{eqnarray}
\citep{mtw:73a}, where $\nu\equiv M_1M_2/(M_1+M_2)^2\le 0.25$ is a symmetrized mass ratio parameter and $r_{gT}\equiv G(M_1+M_2)/c^2$ is the gravitational radius corresponding to the sum of the black hole masses.  Here, we have assumed circular orbits, a natural outcome of both astrophysical processes and gravitational wave emission. We have scaled to a total black hole mass of 60\Msun\ for ease of comparison with LIGO results.  The corresponding time to merger is
\begin{eqnarray}\label{eqn:gw2}
\tau_{\rm GW}&=&\frac{5c^5}{256G^3}\frac{r^4}{M_1M_2(M_1+M_2)}\\
&\approx&2315\,\left(\frac{\nu}{0.25}\right)^{-1}\left(\frac{r}{100r_{gT}}\right)^4\left(\frac{M_1+M_2}{60\Msun}\right)\,\s.
\end{eqnarray}
The frequency of the quadrupolar radiation is twice the orbital frequency, 
\begin{eqnarray}\label{eqn:gw3}
f_{\rm GW}&=&2\left[\frac{G(M_1+M_2)}{r^3}\right]^{1/2}\\
&\approx&6.7\left(\frac{r}{100r_{gT}}\right)^{-3/2}\left(\frac{M_1+M_2}{60\Msun}\right)^{-1}\,\Hz.
\end{eqnarray}
There are important points to note from these expressions.   Firstly, the GW luminosity and frequency rise steeply as the separation decreases ($L_{\rm GW}\propto r^{-5}, f_{\rm GW}\propto r^{-3/2}$).  This gives the inspiral stage a characteristic ``chirp''.  Secondly, noting that $L_{\rm GW}=-GM_1M_2\dot{r}/2r^2$, it can be shown that the  frequency evolution of the inspiral waveform is predominantly governed by the so-called chirp-mass
\begin{equation}
{\cal M}=\frac{(M_1M_2)^{3/5}}{(M_1+M_2)^{1/5}}.
\end{equation}
For a given total mass, the chirp mass and the GW luminosity are maximized, and the time to merger minimized, for equal mass binaries $M_1=M_2$ (in which case the total mass of the binary is given by $2^{6/5}{\cal M}$.  Hence, we expect a selection bias towards seeing equal mass binaries and, as already mentioned, the ten BHBs discovered in LIGO-Virgo O1/O2 are all consistent with being equal mass binaries.  \begin{marginnote}[]
\entry{${\cal M}$}{Chirp mass. A characteristic mass that is most readily determined from the gravitational waveform of a merging BBH.}
\end{marginnote}

As is well known, the two body system in GR does not have a closed form solution.  Equations~\ref{eqn:gw1}-\ref{eqn:gw3}  treat the orbits themselves as Newtonian and then imposes an energy loss at the rate given by the associated quadrupole GW luminosity.  As the black hole separation starts to diminish, the orbital velocities becomes mildly relativistic, and the relativistic corrections to gravity become increasingly important including those that are associated with spin (see Section~\ref{inspiral}).  Even then, the inspiral stage can be very successfully described by analytic approaches such as the effective one body (EOB) formulism \citep{buonanno:99a}.  The inspiral stage ends when the black holes reach the BBH equivalent of the ISCO, after which the black holes plunge together and start their coalescence.

\subsubsection{Coalescence (during)} The final plunge and initial coalescence of the two black holes is an immensely violent process and must be studied by numerical integration of the full non-linear Einstein field equations.  These calculations find that, during the plunge and coalescence, a few percent of the rest mass energy of the system is radiated in a few light crossing times of the event horizon, producing a GW luminosity of $L_{\rm GW}\sim 10^{-3}c^5/G\sim 4\times 10^{56}\ergps$.  For a brief time, the power in GWs from a single merging BBH exceeds the integrated luminosity of every star in the Universe.

Any asymmetry in the final plunge and coalescence due to the spins of the black holes or difference in the component masses ($M_1\ne M_2$) will lead to the asymmetric emission of GWs that will carry away linear momentum.  The final remnant black hole will recoil as a result.   As soon as numerical relativity had progressed to the point where calculations of black hole mergers were possible \citep{pretorius:05a,baker:06a}, these recoil kicks became a focus of study \citep{baker:06b}.  In addition to the mass ratio of the binary, the recoil kick depends sensitively on the magnitude and alignment of the spins of the two black holes.   For non-spinning black holes, the maximum recoil velocity is $175\kmps$ and is achieved for mass differences of $M_1/M_2\approx 1.26$ \citep{gonzales:07a}.  Significantly higher recoil velocities can result if the black holes are spinning.  In the so-called superkick configurations, where the black holes are moderately-to-rapidly spinning with axes that lie in the orbital plane, numerical relativity find that kicks of $2500\kmps$ can be readily obtained, with extreme cases of kicks to $4000\kmps$ possible \citep{campanelli:07a}.  The nature of these kicks will be important in our discussion of possible formation scenarios for the BBHs detected by LIGO.

\subsubsection{Ringdown (after)}The product of coalescence is a rotating black hole with a spacetime that is perturbed with respect to the Kerr metric.  This stage can be treated with linear perturbation theory, and it is found that the dynamics are described by a series of decaying quasi-normal modes as the black hole spacetime relaxes towards the Kerr metric.   The frequencies and decay rates of the oscillation modes are completely determined by the mass and spin of the final black hole.  Thus, if several modes could be detected, one could test the no-hair theorem of GR by seeking the expected consistency relation between the frequencies \citep{isi:19a}. 

\subsection{Spin constraints for the pre-merger black holes}\label{inspiral}

At the current time, essentially all direct GW constraints on black hole spin come from the inspiral stage.  The inspiral waveform is affected by both the magnitude and directions of the black hole spins (measured with respect to the orbital angular momentum).  We begin by introducing the concept of effective spin and precession spin, combinations of spin projections that are most directly imprinted on the inspiral waveform.  We then summarize the spin constraints from the first and second observing runs of LIGO-Virgo.  All ten BBHs catalogued during these two runs were consistent with being equal mass, making it difficult to disambiguate the individual component spins.  However, we end this discussion of inspiral waveforms by addressing the two currently published black hole systems from the third LIGO-Virgo observing run, GW190412 and GW190814, both of which have very unequal masses.  This permits significantly stronger statements to be made about spin.

\subsubsection{Effective and precession spin}\label{gweffspins}

Approximate analytic treatments of the inspiral stage (Post-Newtonian theory or the EOB formulism) show that the phasing of the waveform is sensitive to the mass weighted average of the spins of the two black holes projected onto the direction of the orbital angular momentum, a quantity known as the effective spin,
\begin{equation}\label{eqn:chi_eff}
\chi_{\rm eff}=\frac{M_1\vec{a_1}+M_2\vec{a_2}}{M_1+M_2}\cdot {\hat{L}},
\end{equation}
where $\vec{a_i}$ is the vector pointing along the axis of the $i$th spin with a magnitude $a_i$ equal to the dimensionless spin parameter of the $i$th black hole and $\hat{L}$ is the unit vector pointing in the direction of the orbital angular momentum.  For a generic spin configuration, the waveform phase shifts associated with $\chi_{\rm eff}$ are the dominant signatures associated with black hole spin.  Of course, a measurement of $\chi_{\rm eff}$ alone does not constrain the individual spins. Suppose for example we were to measure $|\chi_{\rm eff}|\ll 1$.  This could result from both black holes having small spins ($|a_1|,|a_2|\ll 1$), both black holes having significant spins in the $\hat{L}$-direction but anti-aligned with each other (so that the contributions cancel out in eqn.~\ref{eqn:chi_eff}), or any significant spin being oriented into the orbital plane.  The exception to this is the case where $\chi_{\rm eff}=\pm 1$ in which case we know that the binary consists of rapidly spinning black holes that are both aligned/anti-aligned with the orbital angular momentum.\begin{marginnote}[]
\entry{$\chi_{\rm eff}$}{BBH effective spin, mass-weighted average of the dimensionless spins projected in the direction of the orbital angular momentum.}
\end{marginnote}

The components of the black hole spins that project into the orbital plane cause precession of the orbital plane.  \cite{schmidt:15a} showed that the precession can be described to leading order by a single parameter, the precession spin $\chi_p$ \citep[the precise expression for $\chi_p$ is slightly more cumbersome than eqn.~\ref{eqn:chi_eff} and hence we simply refer the interested reader to][]{schmidt:15a}.  For generic configurations, precession effects are sub-dominant in the BBH merger waveform and, indeed, only one out of the 12 BBHs reported to date has any evidence for precession (see Section~\ref{gw190412}).\begin{marginnote}[]
\entry{$\chi_p$}{Precession spin of a BBH, a binary averaged quantity related to the components of spins in the orbital plane.}
\end{marginnote}

\subsubsection{Spin results from LIGO-Virgo O1/O2}\label{o1o2results}

\begin{figure}
\includegraphics[width=1.0\textwidth]{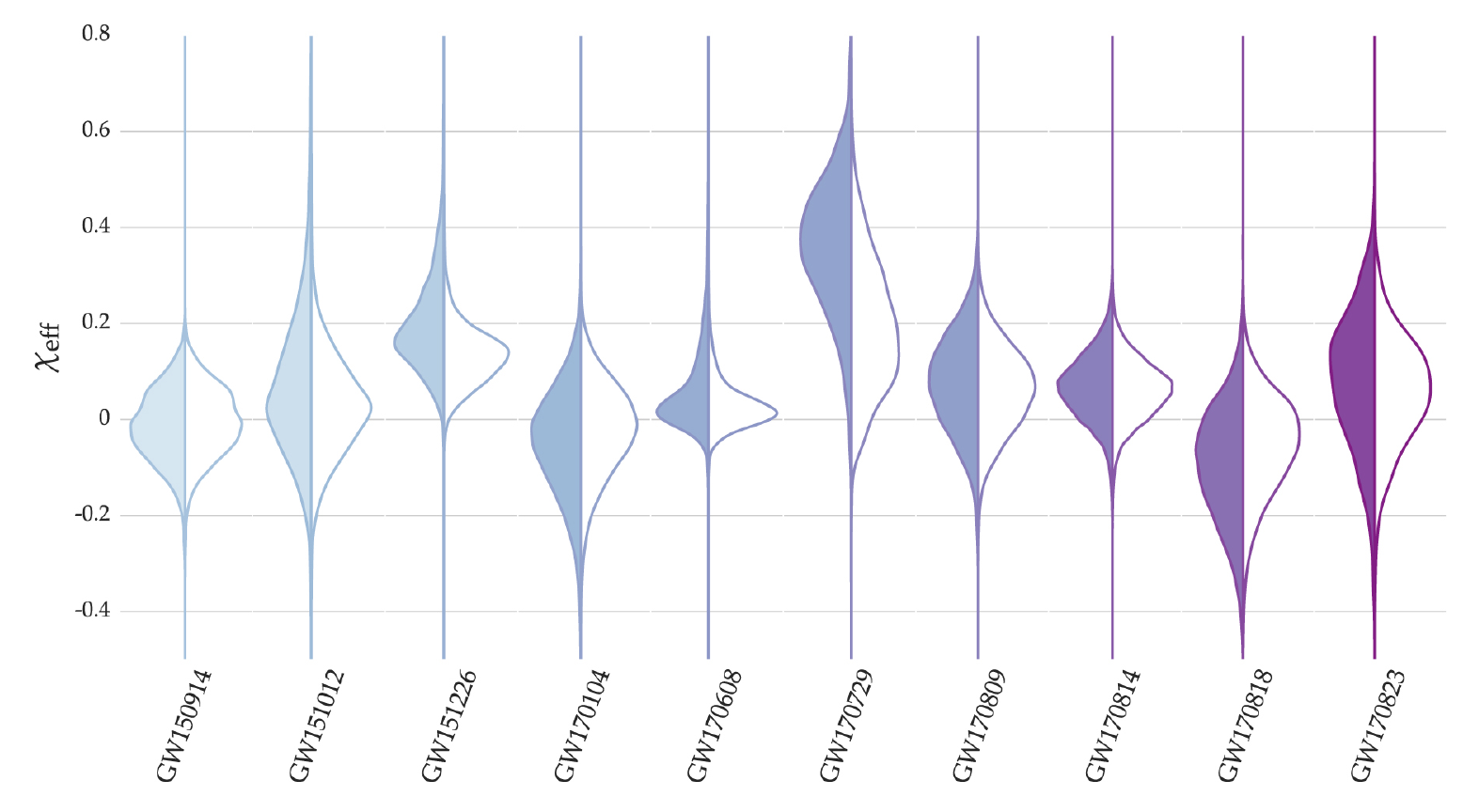}
\caption{The effect of prior assumptions on the derived effective spins $\chi_{\rm eff}$ of the ten BBHs from O1/O2.  The shaded (left) curves show the posterior probabilities using the standard LVC priors, namely uniform priors on the two component masses ($M_1$ and $M_2$), uniform prior on the magnitude of the two dimensionless spins ($a_1$ and $a_2$), and an isotropic prior for the directions of the two spins. Under this analysis, non-zero effective spin is detected at the 90\% confidence level in two events, GW151226 and GW170729.  The unshaded (right) curves show the corresponding posterior probabilities using population-informed priors on $\chi_{\rm eff}$ itself.  The fact that most of the O1/O2 events have very modest or no detectable effective spin suggests that low-$\chi_{\rm eff}$ is a characteristic of the underlying population, so a population informed prior weights the posterior probabilities towards low values of $\chi_{\rm eff}$.   Particularly noteworthy is the case of GW170729 which, with these new priors, no longer requires non-zero effective spin. Figure reproduced from \cite{miller:20a}, \copyright AAS.  Reproduced with permission. }
\label{fig:o1o2effspin_redo}
\end{figure}

The first and second  LIGO-Virgo observing runs achieved compelling detections of ten BBHs \citep[see catalogue presented in][]{abbott:19a}.  These systems have chirp masses between $8\Msun$ and $35-40\Msun$ which, given that they are all consistent with being equal mass binaries, suggests total masses between $18\Msun$ and $80-90\Msun$.  The analysis presented by LVC found evidence for non-zero effective spin at the 90\% confidence level in only two of these events, GW151226 \citep[$\chi_{\rm eff}=0.18^{+0.2}_{-0.12}$;][]{abbott:16b} and GW170729 \citep[$\chi_{\rm eff}=0.37^{+0.21}_{-0.25}$;][]{abbott:19a}.  As can be seen in Fig.~\ref{fig:o1o2effspin_redo} (shaded curves), most of the remaining events are constrained by the LVC analysis to have rather low values of effective spin, with six events possessing $|\chi_{\rm eff}|<0.25$ (90\% confidence level).

These are early days of GW astronomy and, in many of these BBH events, the data are only mildly informative with regards to spin. In this case, the choice of prior can play an important role when posterior probabilities, $p(\chi_{\rm eff})$ and $p(\chi_p)$, are inferred from the data \citep{vitale:17a}. The LVC analyses use priors that are uniform in the two masses ($M_1$ and $M_2$), uniform in the magnitude of the two dimensionless spins ($a_1$ and $a_2$), and isotropic in the directions of the spins.  \cite{miller:20a} have re-analyzed the ten O1/O2 BBHs, arguing the merits of setting priors directly on observable quantities ($\chi_{\rm eff}$) rather than the physical quantities ($a_1, a_2$ and spin directions).  They adopt an astrophysically motivated prior for the two black hole masses and priors on $\chi_{\rm eff}$ that are informed by constraints on the population as a whole.  As shown in Fig.~\ref{fig:o1o2effspin_redo} (unshaded curved), these new priors systematically decrease the magnitude of the inferred $\chi_{\rm eff}$ for each source. Indeed, just by this use of alternative priors, the non-zero spin of GW170729 is no longer significant at the 90\% level.  

Still, beyond simple detection of spin effects, our real interest lies in the spins and spin orientations of the individual black holes.  A number of authors have used the $\chi_{\rm eff}$ constraints to probe the distribution of the magnitudes and directions of black hole spins in the underlying population of detectable BBHs \citep[e.g.][]{farr:17a,tiwari:18a,roulet:19a,miller:20a,biscoveanu:20a}.  \cite{farr:17a} compared the early $\chi_{\rm eff}$ results with expectations from three simple models of the spin probability distribution,
\begin{equation} 
p(a)= 
\begin{cases}
2(1-a) & \mbox{(low spin population)}\\
1 & \mbox{(uniform spin population)}\\
2a & \mbox{(high spin population)}. \\
\end{cases}
\end{equation}
Using just the first four BBH detections, \cite{farr:17a} ruled out the uniform- or high-spin population if it is assumed further that spins are aligned with the orbital angular momentum.  Even in the case of the low spin population, an isotropic distribution of the spins was a better description of the LIGO results than the aligned case.  As the number of sources accumulated, it became increasingly important to account for the degeneracy between $\chi_{\rm eff}$ and the mass ratio, as well as a selection bias that increases the detectability of high effective spin systems (due to the high-$\chi_{\rm eff}$ systems possessing waveforms with more cycles than low-$\chi_{\rm eff}$ systems).  Including these biases and degeneracies in an analysis of the first six BBH detections, \cite{tiwari:18a} found strengthening evidence against aligned spins and, even assuming isotropy, found low-spin populations to be favored over a high-spin populations. 

With the release of the full catalogue of ten BBHs detected in LIGO-Virgo O1/O2, we can go beyond these simple model distributions.  \cite{roulet:19a} employ models for the angular distributions of spin that are uniformly distributed in $\mu=\cos\theta$ across the range $\mu\in[\mu_{\rm min},1]$, providing a series of models that continuously interpolate between an isotropic population ($\mu_{\rm min}=0$) and a perfectly aligned population ($\mu_{\rm min}=1$).  For the spin magnitude distribution, they used a flat distribution across the range $a\in[\bar{a}-0.1,\bar{a}+0.1]$.  Through a full re-analysis of the LIGO-Virgo strain data, \cite{roulet:19a} found that the population must have $\bar{a}<0.4$ for an isotropic spin distribution.  Interestingly, using this more flexible model they find that a perfectly aligned population cannot be ruled out, but then the population must be dominated by slowly spinning black holes,  $\bar{a}<0.1$.  Despite these tight limits on the population-averaged spin, a reanalysis of individual merger events with astrophysically-motivated priors showed that a rapidly spinning black hole was permitted (but not preferred) in four events --- three of these events (GW151012, GW151226, GW170608) are the lowest mass binaries of the O1/O2 set (with chirp masses ${\cal M}<20\Msun$), whereas the fourth event (GW170729) is the most massive (${\cal M}>35\Msun$).

\subsubsection{The asymmetric BBH merger GW190412}\label{gw190412} The first reported event from LIGO-Virgo/O3, GW190412 \citep{abbott:20a}, is remarkable in several ways.  It is the first reported BBH event that is definitively an unequal mass binary, with $M_1=29.7^{+5.0}_{-5.3}\Msun$ and $M_2=8.4^{+1.8}_{-1.0}\Msun$  (90\% credible intervals).  Correspondingly, this event has yielded the first detection of higher-order GW modes (beyond the fundamental quadrupole mode). The inclusion of the higher-order modes in the waveform analysis are important for breaking the degeneracy between the mass ratio and effective spin.  Consequently, this event gives the most significant detection yet of effective spin ($\chi_{\rm eff}=0.25^{+0.08}_{-0.11}$; 90\% confidence range) and the first significant constraints on precessional spin terms ($\chi_p=0.30^{+0.19}_{-0.15}$; 90\% confidence range).  Assuming that the $\chi_{\rm eff}$ is dominated by the more massive black hole, and noting that large values of $\chi_p$ are strongly ruled out (forbidding a significant spin component orthogonal to the orbital angular momentum), the LVC-analysis suggests a primary black hole spin of $a=0.43^{+0.16}_{-0.26}$.

Again, however, in these early days of GW astronomy one must be careful about the effect of priors on astrophysical conclusions.  \cite{mandel:20a} argue that within the context of massive binary star evolution the primary black hole should be born very slowly spinning whereas the secondary black hole should be rapidly spinning and aligned with the binary (due to tidal spin up of its progenitor star by the primary black hole).  Noting that the non-zero detection of precession by the LVC-analysis may not be robust to a different choice of priors, they proceed to neglect precession and re-weight the $\chi_{\rm eff}$-priors of the LVC-analysis to argue that GW190412 is consistent with a non-spinning primary and an aligned rapidly-spinning secondary ($a_2=0.64-0.99$).  \cite{zevin:20a} examines the full strain data with several spin priors (only the primary object spins, only the secondary object spins, both spin, or neither spin), finding that the tentative detection of non-zero precession yields a moderate preference for a spinning primary over models in which only the secondary spins.  

The final conclusion is that this event clearly reveals the presence of black hole spin, but it remains ambiguous whether it was a modestly spinning primary black hole or a rapidly spinning secondary.

\subsubsection{The definitive measurement of low spin in GW190814}\label{gw190814} The LIGO-Virgo/O3 event, GW190814 \citep{abbott:20b}, consists of a 23\Msun\ black hole merging with a 2.6\Msun\ compact object.  The very existence of a secondary of this mass is unexpected, corresponding to either the most massive neutron star or the least  massive black hole yet seen.  From the point of view of black hole spin, this event is remarkable in the precision to which spin effects are absent from the waveform ($\chi_{\rm eff}=-0.002^{+0.060}_{-0.061}$, $\chi_p=0.04^{+0.04}_{-0.03}$; 90\% credible regions).  Given the large mass ratio, these lead to tight constraints on the spin of the primary black hole of $a<0.07$ (90\% confidence level).  To the extent that there are no astrophysical complexities or assumptions needed in this analysis, this is the most robust and constraining measurement of any black hole spin to date.

\subsection{Spin of the remnant black hole}\label{merged}

For the equal mass BBHs that dominate the current known events, the final spin of the remnant black hole ($a_f$) is largely inherited from the orbital angular momentum of the binary, and so is relatively insensitive to the spins or spin orientations of the individual merging black holes.  The final spins reported by LVC \citep{abbott:16a,abbott:19a} derive from fits to numerical relativity simulations of BBH mergers so, strictly, are deduced rather than measured and are always around $a_f\approx 0.7$.  Much of the uncertainty in the final spin results from a degeneracy with the mass-ratio.  

In principle, GWs from the quasi-normal mode ring-down of the final black hole provide a precision probe of the final mass and spin, as well as allowing tests of the no-hair theorem (via consistency relations between the modes).  \cite{isi:19a} report a detection of the fundamental mode and one overtone at $3.6\sigma$ in the LIGO data for GW150914.  They derive a final spin of $a_f=0.63\pm 0.16$ entirely consistent with the $a_f=0.69^{+0.05}_{-0.04}$ reported by LVC \citep{abbott:16a}.  With sufficient sensitivity in the future, ringdown waveforms will enable ``gravitational spectroscopy'', not only enabling precision measurements of the mass and spin of the final black hole, but also testing the no-hair theorem of GW via the consistency of higher order harmonics.

\section{CURRENT RESULTS AND ASTROPHYSICAL IMPLICATIONS}\label{results}

Having described in detail the methodologies for measuring spin, we now discuss the actual results to date and give a flavour of the astrophysical implications.  

\subsection{The spin of SMBHs --- constraints on black hole growth over cosmic time}

\begin{table*}
\begin{center}
\caption{: Summary of published AGN/SMBH spin measurements from the X-ray reflection method in approximate order of increasing mass. Reflecting the conventions in the primary literature, all masses are quoted with $1\sigma$ error bars whereas spins are quoted with 90 per cent error ranges.} 
\begin{tabular}{lcccl}
\hline\noalign{\smallskip}
Object  & Mass ($\times 10^6M_\odot$) & Spin 			& Mass/Spin References 	  \\
\noalign{\smallskip}\hline\noalign{\smallskip}
Mrk359		& $\sim 1.1$ 		& $0.66^{+0.30}_{-0.54}$	& Va16+ \\
Ark564 		& $\sim 1.1$		& $>0.9$				& Va16+/Ji19 \\
Mrk766		& $1.8^{+1.6}_{-1.4}$	& $>0.92$			& Be06/Bu18 \\
NGC4051		& $1.91\pm0.78$	& $>0.99$				&Va16+ \\
NGC1365		& $\sim 2$		& $>0.97$				& Va16+/Wa14 \\
1H0707-495	& $\sim 2.3$		& $>0.94$				& Va16+/Ka15\\
MCG--6-30-15	& $2.9^{+1.8}_{-1.6}$& $0.91^{+0.06}_{-0.07}$	& Va16+/Ma13\\
NGC5506		& $\sim 5$		& $0.93\pm 0.04$		& Ni09/Su18\\
IRAS13224--3809 & $\sim 6.3$		& $>0.975$			& Va16+/Ji18\\
Tons180 		& $\sim 8.1$		& $>0.98$				& Va16+/Ji19 \\
ESO~362--G18 & $12.5\pm 4.5$     & $>0.92$				& VA16+ \\
Swift~J2127.4+5654 & $\sim 15$	& $0.72^{+0.14}_{-0.20}$	& Va16+/Ji19\\
Mrk335  		& $17.8^{+4.6}_{-3.7}$	& $>0.99$			& Gr18/Ji19 	\\
Mrk110		& $25.1\pm 6.1$	& $>0.99$				& Va16+/Ji19\\
NGC3783 	& $29.8\pm 5.4$ 	& $>0.88$ 			& Va16+ \\
1H0323$+$342	& $34^{+9}_{-6}$	& $>0.9$				& Wa16/Gh18 \\
NGC 4151  	& $45.7^{+5.7}_{-4.7}$ & $>0.9$  			& Be06/Ke15 \\
Mrk79		& $52.4\pm 14.4$	& $>0.5$				& Va16+/Ji19\\
PG1229$+$204 & $57\pm 25$		& $0.93^{+0.06}_{-0.02}$	& Ji19/Ji19\\
IRAS13197-1627 & $\sim 64$ 		&  $>0.7$				& Va10/Wa18 \\ 
3C120		& $69^{+31}_{-24}$	& $>0.95$				& Gr18/Va16+ \\
Mrk841		& $\sim 79$		& $>0.52$				& Va16+\\
IRAS09149--6206 & $\sim 100$	& $0.94^{+0.02}_{-0.07}$	& Wa20/Wa20\\
Ark120		& $150\pm 19$		& $>0.85$				& Va16+/Ji19\\
RBS1124		& $\sim 180$		& $>0.8$				& Mi10/Ji19\\
RXS~J1131-1231	& $\sim 200$	& $0.87^{+0.08}_{-0.15}$	& Sl12/Re14c\\
Fairall~9		& $255\pm 56$		& $0.52^{+0.19}_{-0.15}$	& Va16+ \\
1H0419-577 	& $\sim 340$		& $>0.98$				& Va16+/Ji19a\\
PG0804$+$761 & $550\pm 60$	& $>0.97$				& Ji19/Ji19\\
Q2237$+$305	& $\sim 1000$		& $0.74^{+0.06}_{-0.03}$	& Ass11/Re14b\\
PG2112$+$059 & $\sim 1000$		&$>0.83$				& Ve06/Sc10\\
H1821$+$643 & $4500\pm 1500$   & $>0.4$              		& Va16+\\
IRAS~00521--7054 & ---			& $>0.77$				& --/Wa19		\\
IRAS13349$+$2438 &---			& $0.93^{+0.03}_{-0.02}$ 	& --/Pa18\\
Fairall~51		& ---				& $>0.75$				& --/Sv15\\
Mrk~1501		& ---				& $>0.97$				& --/Ch19\\
\noalign{\smallskip}\hline
\end{tabular}\label{tab:smbh}
\end{center}
{\footnotesize Key to references:  
AG14=\cite{agis:14a}, 
Be06=\cite{bentz:06a}, 
Bu18=\cite{buisson:18a}, 
Ch19=\cite{chamani:20a},
Gh18=\cite{ghosh:18a}, 
Gr18=\cite{grier:17a}, 
Ji18=\cite{jiang:18a}, 
Ji19=\cite{jiang:19a}, 
Ka15=\cite{kara:15a}, 
Ke15=\cite{keck:15a}, 
Mi10=\cite{miniutti:10a},
Ni09=\cite{niko:09a}, 
Pa14=\cite{parker:14a}, 
Pa18=\cite{parker:18a}, 
Re14b=\cite{reynoldsm:14a}, 
Re14c=\cite{reis:14a}, 
Va10=\cite{vasudevan:10a},
Va16+=\cite{vasudevan:16a} and references there within,
Ve06=\cite{vestergaard:06a} 
 Sc10=\cite{schartel:10a}, 
 Su18=\cite{sun:18a}, 
 Sv15=\cite{svoboda:15a}, 
 Wa16=\cite{wang:16a}, 
 Wa18=\cite{walton:18a}, 
 Wa19=\cite{walton:19a}, 
  Wa19=\cite{walton:20a}.
 }
\end{table*}

\begin{figure}
\includegraphics[width=1.0\textwidth]{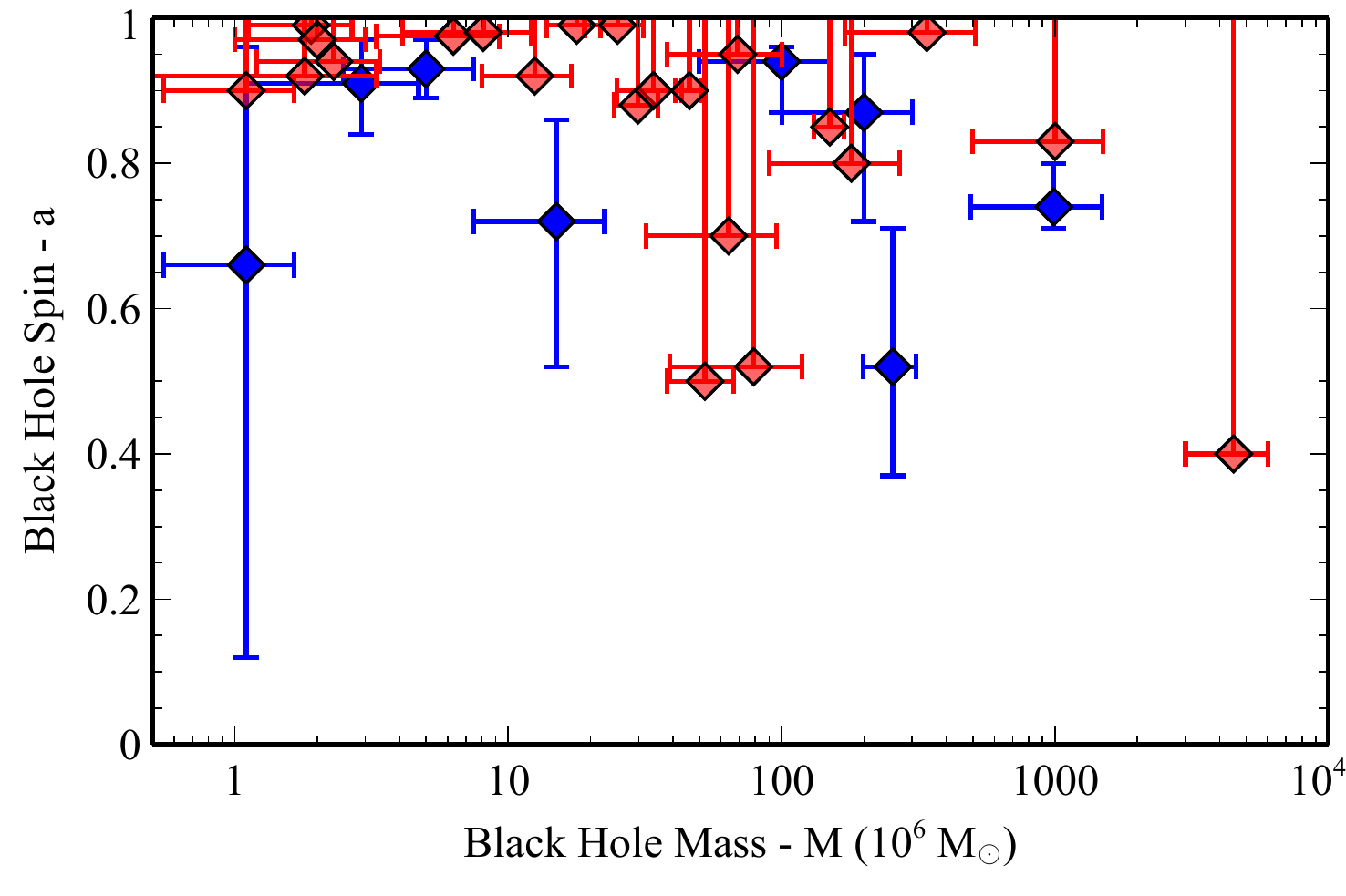}
\caption{SMBH spins as a function of mass for the 32 objects in Table~\ref{tab:smbh} that have available mass estimators.  All spin measurements reported here are from the X-ray reflection method.  Lower limits are reported in red, and measurements that include a meaningful upper bound (distinct from $a=1$) are reported in blue.   Following the convention of the relevant primary literature, error bars in spin show the 90\% confidence range.  The error bars in mass are the $1\sigma$-errors from Table~\ref{tab:smbh} or, where that is not available, we assume a $\pm 50$\% error.}
\label{fig:smbh_spins}
\end{figure}

For all of the reasons outlined in Section~\ref{accreting}, robust individual-object constraints on the spin of SMBHs are dominated by the X-ray reflection method.  Table~\ref{tab:smbh} and Fig.~\ref{fig:smbh_spins} summarize the current reflection-based AGN spin measurements; this updates the compilations of \cite{reynolds:14a} and \cite{vasudevan:16a} applying the same ``quality control criteria'' and, in particular, includes the results obtained from the recent high-density reflection modelling of \cite{jiang:19a}.  Taken together, these results suggest a mass dependence to the distribution of SMBH spins.  Most SMBHs with masses $M<3\times 10^7\Msun$ are found to be spinning very rapidly, $a>0.9$, whereas a population of more modestly spinning black holes start to emerge at higher masses.  

There are a few caveats that must be attached to these results.  Firstly, while all of these spins have been derived from the X-ray reflection method, these works have been conducted by many independent groups and there is a heterogeneous methodology for describing the other spectral complexity at play in these data.  In most cases, only statistical errors were quoted and systematic effects resulting from, for example, finite disk thickness effects (which could shift results either way by $\Delta a\approx 0.1$) are not included.  Secondly, the more massive black holes in this sample are found in the more distant quasars which typically fainter; indeed, measurements were only possible in Q2237$+$305 \citep{reynoldsm:14a} and RXSJ1131--1231 \citep{reis:14a} by utilizing strong gravitational lensing.  Limited data quality for these more massive SMBHs will lead to weaker limits on spin and hence error bars that extend to smaller spin values. Still, we find that even if we restrict limit ourselves to spin measurements with a meaningful upper-bound (blue points on Fig.~\ref{fig:smbh_spins}), hints of this mass dependence survive.  Finally, these results should not be taken as fully representative of the SMBH population as there is an important selection bias at play.  The 34 objects listed in Table~\ref{tab:smbh} were singled out for study in part because their were sufficiently X-ray bright to obtain high-quality X-ray spectra.  Given that the radiative efficiency of black hole accretion is a strong function of black hole spin, rapidly spinning black holes are expected to be more luminous for a given mass accretion rate and so over represented in any flux limited sample \citep{brenneman:11a,vasudevan:16a}.

These caveats aside, the observed mass-dependent spin distribution is an important constraint on models of SMBH growth.  The predominance of rapid-spinners in the low-mass SMBH population suggests growth via coherent accretion events, whereas the more modest spins of the more massive SMBHs ($M>3\times 10^7\Msun$) suggests the growing importance of SMBH-SMBH major mergers and/or incoherent accretion \citep{volonteri:05a}.  Indeed, this exact mass dependence of SMBH-spins is seen in semi-analytic \citep{sesana:14a,zhang:19a} and full cosmological-hydrodynamic \citep{bustamante:19a} models of hierarchal galaxy formation.

\subsection{The spin of stellar mass black holes --- the need for multiple populations}\label{stellarform}

\begin{table}
\caption{Measurements of black hole spin in dynamically-confirmed black holes X-ray binaries from the X-ray reflection and CF-method.   Spins are listed with their 90\% error ranges.  }
\label{tab:xrbspins}       
\begin{center}
\begin{tabular}{lccl}
\hline\noalign{\smallskip}
Object  			& Spin from Reflection 		& Spin from CF 		& References  \\
\noalign{\smallskip}\hline\noalign{\smallskip}
IC~10 X-1		&	 ---					& $0.85^{+0.04}_{-0.07}$		& --/St16\\
M33 X-7			&	---					& $0.84\pm 0.05$			& --Li08\\
LMC X~3			&	 ---					& $0.25^{+0.20}_{=0.29}$		& --/St14\\
LMC X~1			& $>0.55$					& $0.92^{+0.05}_{-0.07}$		& St12/Tr20\\
AO620--00		&	 --- 					& $0.12\pm 0.19$			& --/Go10\\
Nova Mus 1991		&	--- 					& $0.63^{+0.16}_{-0.19}$		& --/Ch16\\
GS~1354--645		& $>0.98$					& ---						&EB16/--\\				
4U~1543--475 		& $0.67^{+0.15}_{-0.08}$		& $0.8\pm 0.1$				& Do20/Sh06\\
XTE~J1550--564 	&	$0.33-0.77$ 			& $0.34^{+0.37}_{-0.45}$ 		& Mi09/St11\\
4U~1630--472		&	$>0.97$				& ---					& Ki14\\
XTE~J1650--500 	&	$0.79\pm 0.01$			& ---					& Mi09/--\\
XTE~J1652--453 	&	$0.45\pm 0.02$			& ---					& Hi11/-- \\
GRO~J1655--40 	&	 $>0.9^*$				& $0.7\pm 0.1$ 			& Rei09/Sh06\\
GX339--4 			& $>0.95$					& ---						& Ji19/--\\
SAX~J1711.6--3808 &	$0.6^{+0.2}_{-0.4}$ 	& ---					& Mi09/--\\
GRS1716--249		&	\multicolumn{2}{c}{$a>0.92$ (joint CF/reflection fit)} 			& Ta19\\
XTE~J1752--223 	& $0.92\pm 0.06$			& ---						& Ga18/--\\
Swift~J1753.5--0127 &	$0.76^{+0.11}_{-0.15}$ & ---					& Rei09/--\\
MAXI~J1836--194 	&	 $0.88\pm 0.03$		& ---					& Rei12/-- \\
EXO~1846--031	& $>0.99$					& ---					& Dr20/--\\
XTE~J1908$+$094 &	 $0.75\pm 0.09$		& ---					& Mi09/--\\
Swift~J1910.2--0546	&	$<-0.32$				& --- 					& Re13\\
GRS1915$+$105 	&	$0.88^{+0.06}_{-0.13}$	& $>0.95$				& Sh20/Mc06\\
V404 Cyg			&	$>0.92$				& ---					& Wa17/-- \\
\noalign{\smallskip}\hline
\end{tabular}
\end{center}
{\footnotesize Key to reference : Ch16=\cite{chen:16a}, Do20=\cite{dong:20a}, Dr20=\cite{draghis:20a}, EB16=\cite{elbatal:16a}, Fa12=\cite{fabian:12b}, Ga18=\cite{garcia:18a}, Go11=\cite{gou:11a}, Go10=\cite{gou:10a}, Hi11=\cite{hiemstra:11a}, Ki14=\cite{king:14a}, LI08=\cite{liu:08a}, Mc06=\cite{mcclintock:06a}, Mi09=\cite{miller:09a}, Mo14=\cite{morningstar:14a}, Rei09=\cite{reis:09a}, Rei12=\cite{reis:12a}, Rei13=\cite{reis:13a}, Sh06=\cite{shafee:06a}, Sh20=\cite{shreeram:20a}, St11=\cite{steiner:11a}, St12=\cite{steiner:12a}, St14=\cite{steiner:14a}, St16=\cite{steiner:16a}, Ta19=\cite{tao:19a}, Tr20=\cite{tripathi:20a}, Wa17=\cite{walton:17a}}
\label{tab:xrbspin}
\end{table}

Table~\ref{tab:xrbspins} reports the X-ray reflection and CF derived spins for the black holes in X-ray binaries.  As in the case of SMBHs, these measurements have been obtained by a number of independent groups using methodologies that differ in detail (observatories and corresponding X-ray band-pass, selection of data and/or spectral state, and methods for marginalizing over unknown or poorly known quantities).   Still, there are several points to comment upon.  Firstly, some of these black hole X-ray binaries provide an opportunity to measure spins with both the X-ray reflection and CF-methods.  In general, there is very good agreement, the exception being a modest tension for GRO~J1655--40.   Secondly, there are a large number of rapidly spinning objects.  Interestingly, the efficiency-spin bias noted in the context of SMBHs should be less relevant here as the black hole X-ray binary sample comes close to being volume limited (mostly consisting of all bright persistent systems in our Galaxy, and all transient systems in our Galaxy that have undergone outburst since the deployment of all sky monitors starting with RXTE in 1995).  

An important question is whether these black holes were born rapidly spinning, or were they spun up by accretion. For a black hole to spin up from $a=0$ to $a\approx 1$ requires it to accrete a significant fraction of its original mass from a rotationally supported accretion disk.  In low-mass black hole X-ray binaries (LMXBs, where the companion is a long-lived low-mass $\sim 1\Msun$ star that fills the Roche Lobe), spin up to moderately rapid spins may just be possible.  \cite{fragos:15a} present detailed binary mass-transfer scenarios that successfully reproduce the high spin of observed LMXBs, spinning up black holes that are born with negligible spin.  An important caveat to these models, however, is that they assume conservative mass transfer.  LMXBs are observed to possess powerful rotating winds from the inner accretion disk \citep{miller:15b}, and it remains unclear whether the spin-up scenario is still viable given the corresponding mass losses.  

The situation is clearer for high-mass X-ray binaries (HMXBs).  The accretion process is time limited due to the rapid evolution of the massive companion star.  Even accreting at the Eddington limit, there would be insufficient time for the black hole to spin up before the companion star undergoes a core-collapse supernova.  Thus, the rapidly spinning black holes seen in many high-mass X-ray binaries had to be born rapidly spinning, providing a window into the structure of the pre-explosion star and the dynamics of the core collapse that produces the black hole.  In the absence of these observational constraints, one might have expected the core-collapse of a massive star in a close binary with another massive star to produce a slowly spinning black hole due to tidal locking of the stellar interiors with the orbit.  The fact that such systems can possess rapidly rotating black holes may suggest very poor mixing of angular momentum during the post main-sequence evolution of the black hole progenitor \citep{qin:19a} and/or a failed supernova followed by the late-time fall back of tidally spun-up debris \citep{batta:17a}.

By stark contrast, the pre-merger black holes of the BBHs seen in GWs are characterized by low spins \citep{farr:17a,tiwari:18a,abbott:19a,miller:20a,abbott:20b}.  While the presence of one rapidly spinning component is consistent with LIGO-Virgo data in a small number of the GW events \citep[particularly GW151226, GW170729, and maybe the low-mass secondary of GW190412; ][]{roulet:19a,mandel:20a}, for none of these systems is it the preferred model.  An analysis of the full set of O1/O2 LIGO-Virgo BBHs accounting for selection effects and biases show the population of pre-merger black holes to have $\bar{a}<0.4$ assuming isotropic spins, with tighter limits $\bar{a}<0.1$ if we adopt as a prior that the spins are aligned with the orbit \citep{roulet:19a}.  

These spin results clearly highlight the need for a diversity of black hole formation channels \citep{draghis:20a,miller:20a}, with the black holes in close BBHs coming from distinct routes to those in HMXBs.  Of course, another important observation is that many of the pre-merger black holes seen to date by GW observatories are appreciably more massive ($M\sim 20-40\Msun$) than those seen in X-ray binaries \citep[$M\sim 10\Msun$][]{orosz:02a}.  In order to form a pair of such massive black holes via regular stellar evolution requires a binary consisting of two particularly massive stars, as well as low-metallicity so that the stellar winds are weak and do not remove appreciable mass.  

An alternative is BBH formation via dynamical processes in a dense stellar cluster.  In this picture, $\sim 10\Msun$ black holes are born via isolated stellar evolution, sink to the center of the cluster due to dynamical friction, and then start to capture each other into BBHs \citep{millerc:02a}.  Indeed, multiple merger generations may be present in events already observed.  GW190412 is an interesting case in point, consisting of a $\sim 8\Msun$ black hole merging with a $\sim 30\Msun$ black hole (see Section~\ref{gw190412}).  \cite{rodriguez:20a} suggest that this is a 3rd generation merger, with the $30\Msun$ black hole resulting from two prior merging events.  As discussed in Section~\ref{gw190412}, this event shows strong evidence for non-zero effective spin and tentative evidence for non-zero precession spin (Section~\ref{gweffspins}), indicating that the more massive black hole possesses a spin of $a=0.43^{+0.16}_{-0.26}$.  On the face of it, we might have expected the more massive black hole to possess a higher spin if it is indeed a 2nd generation merger product itself.  However, the observed spin of the primary is consistent with the 3rd generation merger scenario once we apply the additional constraint that the merger product is retained in the stellar cluster even after the GW-induced recoil of the second merger \citep{rodriguez:20a}.  Turning this argument around, the extremely slow-spin of the 23\Msun\ black hole in GW190814 argues against this object being itself a merger remnant. 

This is far from an exhaustive discussion of the many channels by which these stellar-mass black holes can form --- it is a rich and rapidly-developing subject deserving of its own dedicate review \citep[see][for a discussion informed by early LIGO results]{mirabel:17a}.  In addition to black holes resulting from the core-collapse of massive stars, there are more exotic channels that are being actively considered such as primordial black hole (PBH) formation during cosmological phase transitions in the early Universe \citep{zeldovich:66a} or the PBH induced collapse of a neutron star \citep{tsai:20a}, and black hole spin remains an important constraint on such models \citep{mirbabayi:20a}. However, further discussion of these more unusual channels of black hole formation is beyond the scope of this review.

\section{CONCLUSIONS AND BRIEF LOOK TO THE FUTURE}\label{future}

The study of black hole spin has rapidly moved from the realm of the theoretical to one driven by an increasingly diverse set of observations. In this review, we have highlighted the principal ways that black hole spin is measured using today's EM and GW observatories, tying back to the basic physics of the black hole and accretion flow where possible.  Quantitative measures of spin from EM observations of accreting black holes have focused on systems in the Goldilocks zone of accretion rate, $\sim 0.01-0.3\dot{M}_{\rm Edd}$, where the accretion disk is geometrically-thin, optically-thick and radiatively-efficient.  Here, the spin-dependent ISCO can impose a clean transition on the accretion flow, leading to spin information being imprinted on the X-ray reflection spectrum (Section~\ref{reflection}), thermal continuum (Section~\ref{continuum}), and radiative-efficiency (Section~\ref{efficiency}).  Given the clear cut assumptions used to extract black hole spin from X-ray reflection signatures or the thermal continuum, one can examine reasonable deviations from those assumptions and hence assess systematic errors.  

Outside of this optimal accretion rate rate, the imprint of spin on the observables becomes more dependent on less established models.  Still, spin can be imprinted on the frequencies of quasi-periodic oscillations (Section~\ref{qpos}) and the total jet power (Section~\ref{jet}).  Furthermore, the ability to image the central regions of SMBH accretion flows either indirectly via the microlensing of gravitationally-lensed quasars (Section~\ref{microlensing}) or directly via mm-band VLBI (Section~\ref{imaging}) allows unique spin probes in a few special systems.  

The results from all of these techniques show that the observed population of accreting black holes, both SMBHs and the black hole X-ray binaries, is dominated by rapidly spinning objects, $a>0.9$.  In the SMBH realm, we find evidence for a population of more modestly spinning black holes at the highest masses ($M>3\times 10^7\Msun$), in line with models that suggest SMBH-SMBH mergers and incoherent accretion should dominate at the highest SMBH masses.  In the case of black hole X-ray binaries, the evidence suggests that many of the black holes must have been born rapidly spinning.  This constrains the structure of the pre-collapse star and the dynamics of the supernova explosion.

The advent of GW astronomy has opened a new window on black holes that is unaffected by the complexities of accretion physics (Section~\ref{gwa}).  At the current time, GW observations are most sensitive to two specific binary-averaged spins, the effective spin relating to spins aligned with the orbital angular momentum and the precession spin relating to spins that lie in the orbital plane (Section~\ref{gweffspins}).  Given that most of the BBH events seen to date are approximately equal mass systems, this leaves the spins and spin directions of the individual pre-merger black holes largely undetermined.  Furthermore, at the current time, the data are often only mildly informative about spin, so conclusions about even the effective and precession spin of the binary is influenced by the assumed priors.  But robust signatures of spin are detected in at least two BBH events, GW151226 and GW190412 (Section~\ref{gw190412}), and there are a small number of events for which the waveform permits one of both of the black holes to be rapidly spinning.  

Still, it is very clear that the bulk of the population of the observed BBH merger events involve slowly spinning black holes, a conclusion that is strengthened by the remarkably strong spin constraint ($a<0.07$) on the $23\Msun$ primary black hole in GW190814 (Section~\ref{gw190814}).  So, comparing the GW view of black holes with those seen in black hole X-ray binaries, spin reveals a stark difference in the two populations and the need for a diversity of formation paths (Section~\ref{stellarform}).  

The future of black hole spin as an observational science is as diverse as it is bright.  The next generation of high-throughput X-ray observatories (e.g. {\it XRISM, Athena, Lynx}, and {\it Strobe-X}) will improve our understanding of accreting black holes, allowing more precise and accurate spin measurements from X-ray reflection spectroscopy, X-ray iron line reverberation, thermal continuum fitting, and QPOs. As well as deepening our knowledge of the local systems that dominate today's studies, these future observatories will permit the characterization of SMBH spin in high-redshift AGN, providing new constraints on the growth of SMBHs over cosmic time.  Continued developments in mm-band and sub-mm VLBI, including the deployment of space-based antennae, will dramatically improve our understanding of ergospheric-scale structures in the closest SMBHs with direct implications for spin determination.  For the closest SMBH in our Galactic Center, there is the real prospect of detecting the Lens-Thirring precession of test-particle orbits if we can find a radio pulsar in a close orbit, an important goal of the Square Kilometer Array (SKA).  Finally, GW astronomy is poised for an explosion of activity, starting with further developments of the ground-based capabilities (improving the sensitivity and spatial location ability of the 30--300\Hz\ GWs and extending the reach down in frequencies to $\sim 1\Hz$ with underground facilities and new technologies) that will permit precision studies of the BBHs discussed in Section~\ref{gwa}.  Ultimately, the deployment of space-based GW detectors (e.g. {\it LISA}) will allow the sensitive characterization of the $\sim 10^{-3}\Hz$ GWs from SMBH-SMBH mergers out to the redshift of SMBH formation, as well as precision mapping of the Kerr metric as smaller black holes chaotically spiral into nearby SMBHs.

\section*{DISCLOSURE STATEMENT}
The authors are not aware of any affiliations, memberships, funding, or financial holdings that
might be perceived as affecting the objectivity of this review. 

\section*{ACKNOWLEDGMENTS}
The author thanks useful and insightful discussions with Philip Armitage, Andy Fabian, Neil Cornish, Erin Kara, Corbin Taylor, and Dom Walton.  The author thanks the UK Science and Technology Facilities Council (STFC) for support under the New Applicant grant ST/R000867/1, and the European Research Council (ERC) for support under the European Union?s Horizon 2020 research and innovation programme (grant 834203).

%

\bibliographystyle{ar-style2}

\end{document}